\pdfoutput=1

\documentclass[11pt,twoside,a4paper,cmspaper,final,collab]{cms-tdr}
\def\svnVersion{285201}\def\cmsCernNo{2013-037}\def\cmsCernDate{\today}\def\cmsMessage{Published in the Journal of High Energy Physics as \href{http://dx.doi.org/10.1007/JHEP04(2015)025}{\doi{10.1007/JHEP04(2015)025.}}}

\begin{document}\cmsNoteHeader{EXO-12-061}

\hyphenation{had-ron-i-za-tion}
\hyphenation{cal-or-i-me-ter}
\hyphenation{de-vices}
\RCS$Revision: 277883 $
\RCS$HeadURL: svn+ssh://alverson@svn.cern.ch/reps/tdr2/papers/EXO-12-061/trunk/EXO-12-061.tex $
\RCS$Id: EXO-12-061.tex 277883 2015-02-17 21:37:37Z alverson $

\newlength\cmsFigWidth
\ifthenelse{\boolean{cms@external}}{\setlength\cmsFigWidth{0.7\textwidth}}{\setlength\cmsFigWidth{0.7\textwidth}}
\ifthenelse{\boolean{cms@external}}{\providecommand{\cmsLeft}{top}}{\providecommand{\cmsLeft}{left}}
\ifthenelse{\boolean{cms@external}}{\providecommand{\cmsRight}{bottom}}{\providecommand{\cmsRight}{right}}
\cmsNoteHeader{EXO-12-061}

\newcommand{\anaLumimm}{20.6}
\newcommand{\anaLumiee}{19.7}

\newcommand{\SSMwidth}{80}
\newcommand{\Esixwidth}{14}
\newcommand{\Gonewidth}{35}
\newcommand{\Gtwowidth}{9.0}
\newcommand{\Gthreewidth}{3.5}

\newcommand{\limitmumuZssm}{2.73}
\newcommand{\limitmumuZpsi}{2.39}
\newcommand{\limitmumuGadd}{1.13}
\newcommand{\limitmumuGlow}{2.12}
\newcommand{\limitmumuGhigh}{2.56}
\newcommand{\limiteeZssm}{2.67}
\newcommand{\limiteeZpsi}{2.34}
\newcommand{\limiteeGadd}{1.25}
\newcommand{\limiteeGlow}{2.13}
\newcommand{\limiteeGhigh}{2.50}
\newcommand{\limitZssm}{2.90}
\newcommand{\limitZpsi}{2.57}
\newcommand{\limitGadd}{1.27}
\newcommand{\limitGlow}{2.35}
\newcommand{\limitGhigh}{2.73}

\newcommand{\cPZg}{\ensuremath{\cmsSymbolFace{Z}/\gamma^{*}}\xspace}
\newcommand{\zmm}{\ensuremath{\cmsSymbolFace{Z}/\gamma^{*}\to \Pgmp\Pgmm}\xspace}
\newcommand{\zee}{\ensuremath{\cmsSymbolFace{Z}/\gamma^{*}\to \Pep\Pem}\xspace}
\newcommand{\ztt}{\ensuremath{\cmsSymbolFace{Z}/\gamma^{*}\to \Pgt^+\Pgt^-}\xspace}

\newcommand{\GKK}{\ensuremath{\mathrm{G}_\mathrm{KK}}\xspace}
\newcommand{\ZPSSM}{\ensuremath{\cPZpr_\mathrm{SSM}}\xspace}
\newcommand{\ZPPSI}{\ensuremath{\cPZpr_\psi}\xspace}

\newcommand{\LT}{\ensuremath{\Lambda_{\mathrm{T}}}\xspace}
\newcommand{\MS}{\ensuremath{M_{\mathrm{S}}}\xspace}
\newcommand{\LC}{\ensuremath{\Lambda}\xspace}

\newcommand{\Mmax}   {\ensuremath{M_{\text{max}}}\xspace}
\newcommand{\mmin}   {\ensuremath{m^{\text{min}}}\xspace}
\newcommand{\mminmm} {\ensuremath{m^{\text{min}}_{\mu\mu}}\xspace}
\newcommand{\mminee} {\ensuremath{m^{\text{min}}_{\Pe\Pe}}\xspace}
\newcommand{\mminll} {\ensuremath{m^{\text{min}}_{\ell\ell}}\xspace}

\title{Search for physics beyond the standard model in dilepton mass spectra in proton-proton collisions at $\sqrt{s} = 8\TeV$}
\titlerunning{BSM physics in dilepton mass spectra}

\date{2013/02/25}

\abstract{
  Dimuon and dielectron mass spectra, obtained from data resulting
  from proton-proton collisions at 8\TeV and recorded by the CMS
  experiment, are used to search for both narrow resonances and broad
  deviations from standard model predictions. The data correspond to
  an integrated luminosity of $\anaLumimm$ ($\anaLumiee$)\fbinv for
  the dimuon (dielectron) channel. No evidence for non-standard-model
  physics is observed and 95\% confidence level limits are set on
  parameters from a number of new physics models.  The narrow
  resonance analyses exclude a Sequential Standard Model $\ZPSSM$
  resonance lighter than $\limitZssm\TeV$, a superstring-inspired
  $\ZPPSI$ lighter than $\limitZpsi\TeV$, and Randall--Sundrum
  Kaluza--Klein gravitons with masses below $\limitGhigh$,
  $\limitGlow$, and $\limitGadd\TeV$ for couplings of 0.10, 0.05, and
  0.01, respectively.  A notable feature is that the limits have been
  calculated in a model-independent way to enable straightforward
  reinterpretation in any model predicting a resonance structure. The
  observed events are also interpreted within the framework of two
  non-resonant analyses: one based on a large extra dimensions model
  and one based on a quark and lepton compositeness model with a
  left-left isoscalar contact interaction. Lower limits are
  established on $\MS$, the scale characterizing the onset of quantum
  gravity, which range from 4.9 to 3.3\TeV, where the number of
  additional spatial dimensions varies from 3 to 7.  Similarly, lower
  limits on $\LC$, the energy scale parameter for the contact
  interaction, are found to be 12.0 (15.2)\TeV for destructive
  (constructive) interference in the dimuon channel and 13.5
  (18.3)\TeV in the dielectron channel.}

\hypersetup{%
pdfauthor={CMS Collaboration},%
pdftitle={Search for physics beyond the standard model in dilepton mass spectra in proton-proton collisions at sqrt(s) = 8 TeV},%
pdfsubject={CMS},%
pdfkeywords={CMS, dileptons, narrow resonances, extra dimensions, compositeness, contact interactions}}

\maketitle

\section{Introduction}

This paper describes a general investigation for evidence of physics
beyond the standard model (SM) using the dilepton (dimuon and
dielectron) invariant mass spectra obtained from $\sqrt{s}=8\TeV$
proton-proton (pp) collision data collected by the CMS detector at the
CERN LHC~\cite{lhc}. The analyses include searches both for new
narrow resonances and for deviations from SM expectations at high
invariant mass values that do not result in a resonance structure.

Numerous new physics models predict the existence of narrow resonances
at the \TeVns{} mass scale. The approach described in this paper is
designed to be independent of specific model assumptions, allowing the
results to be reinterpreted for any model predicting a spin-1 or
spin-2 narrow resonance.  A generic resonance is denoted by $\cPZpr$
in this paper; wherever a specific model is implied a subscript is
used to specify the model.  The search results are interpreted in the
context of various models: the Sequential Standard Model $\ZPSSM$ with
SM-like couplings~\cite{Altar:1989}, the $\ZPPSI$ possible in grand
unified theories where the gauge group is
E$_6$~\cite{Leike:1998wr,Zp_PSI_3}, and Kaluza--Klein graviton
($\GKK$) excitations arising in the Randall--Sundrum (RS) model of
extra dimensions~\cite{Randall:1999vf, Randall:1999ee}.  For a
resonance mass of 2.5\TeV, the widths of the $\ZPSSM$ and $\ZPPSI$ are
\SSMwidth\ and \Esixwidth\GeV. Similarly, the $\GKK$ widths are
\Gthreewidth, \Gtwowidth, and \Gonewidth\GeV for $\GKK$ coupling
parameters $k/\overline{M}_\mathrm{Pl}$ of 0.01, 0.05, and 0.10, where
$k$ is the warp factor of $n$-dimensional anti-de Sitter space and
$\overline{M}_\mathrm{Pl}$ is the reduced Planck scale.

 Non-resonant deviations from the SM are interpreted within two
frameworks: the (1) Arkani-Hamed--Dimopoulos--Dvali (ADD)
model \cite{arkani98:hlz, arkani99:hlz}, where possible enhancements
in the high invariant mass cross section are due to virtual
graviton-mediated processes, and (2) contact interactions,
specifically the left-left isoscalar model~\cite{TheoryII}, where
possible deviations are due to quark and lepton substructures.

In the ADD model, three possible parametrizations of the differential
cross section are given by Hewett \cite{Hewett1999}, Han, Lykken,
Zhang (HLZ) \cite{han99:hlz}, and Giudice, Rattazzi, Wells (GRW)
\cite{GRW99:extradim}.  In the HLZ convention, the parameters that
define the characteristics of the model are the ultraviolet cutoff
scale of the divergent sum of the KK graviton excitations
$\MS$, which characterizes the onset of quantum gravity
effects~\cite{Franceschini2011}, and the number of extra spatial
dimensions $n_{\mathrm{ED}}$. These can be related to the
similar ultraviolet cutoff scale $\LT$ in the GRW convention:
\begin{equation}
\LT^{-4} =\begin{cases}
 \MS^{-4} \log\left(\frac{\MS^2}{M_{\ell\ell}^2}\right), &n_{\mathrm{ED}}= 2;\\
 \frac{2}{n_{\mathrm {ED}}-2} \MS^{-4}, &n_{\mathrm{ED}} > 2,\end{cases}
\end{equation}
where $M_{\ell\ell}$ is the dilepton invariant mass. The HLZ and GRW
conventions are based on effective field theory, which is expected to
break down at the energy scale $\Mmax$ at which quantum gravity
effects become strong, \ie close to $\MS$ (HLZ) or $\LT$ (GRW). In
this analysis, when presenting results in the HLZ (GRW) convention it
is assumed that $\Mmax$ is equal to $\MS$ ($\LT$).  The results do not
depend strongly on the exact choice of $\Mmax$ for a broad range of
$\Mmax$ values around the chosen cutoff
scale~\cite{CMS2012:ADDdilepton}.

The data are also interpreted in the context of possible contact
interactions (CI).  The existence of three families of quarks and
leptons might be explained if these particles are composed of more
fundamental constituents. In order to confine the constituents and to
account for the properties of quarks and leptons, a new strong gauge
interaction, metacolor, is introduced. Below a given interaction
energy scale $\LC$, the effect of the metacolor interaction is to bind
the constituents into metacolor-singlet states.  For parton-parton
centre-of-mass energies less than $\LC$, the metacolor force will
manifest itself in the form of a flavour-diagonal CI~\cite{TheoryI}.
The model considered here is the left-left isoscalar model, which is
the conventional benchmark for CI in the dilepton channel. The
predicted differential cross section includes an interference term,
which may be positive or negative, and the measurements are
interpreted in the context of both possibilities.

Results of searches for narrow $\zp\to{\ell^+\ell^-}$
($\ell=\mu,\Pe$) resonances in pp collision data have previously
been reported by the ATLAS and CMS
Collaborations~\cite{CMS-dilep-2012,ATLAS-dilep-2014}.  The CDF and D0
Collaborations have published results based on a $\Pp\Pap$ collision
sample at $\sqrt{s} = 1.96\TeV$ and ${\approx}$5\fbinv of integrated
luminosity~\cite{CDF_Zp,CDF_RS,D0_RS,D0_Zp,CDF_SSM,CDF_RSele}.
Similarly, there are recent limits from the LHC on the
ADD~\cite{CMS2012:ADDdilepton,ATLASCIADD2014} and
CI~\cite{ATLAS2,ATLAS3,PhysRevD.87.015010,CMS_ci_dimuon_7_TEV,ATLASCIADD2014}
model parameters in dilepton channels.

The results presented in this paper are obtained from an analysis of a
data sample collected in 2012 at $\sqrt{s}=8\TeV$, corresponding to an
integrated luminosity of $\anaLumimm$ ($\anaLumiee$)\fbinv for the
dimuon (dielectron) channel. The data used have been processed with
the most recent calibration and alignment constants for all detector
elements.

\section{The CMS detector}\label{sec:CMS}

The central feature of the CMS detector is a superconducting solenoid
providing an axial magnetic field of 3.8\unit{T} and enclosing the
all-silicon inner tracker, the crystal electromagnetic calorimeter
(ECAL), and the brass and scintillator hadron calorimeter (HCAL). The
inner tracker is composed of a silicon pixel detector and a silicon
strip tracker, and measures charged-particle trajectories in the
pseudorapidity range $\abs{\eta}<2.5$.
The finely segmented ECAL consists of nearly 76\,000 lead tungstate
crystals, which provide coverage in pseudorapidity up to $\abs{\eta} =
3.0$.  The muon system covers the pseudorapidity region $\abs{\eta}<2.4$
and consists of up to four stations of gas-ionization muon detectors
installed outside the solenoid and sandwiched between the layers of
the steel flux-return yoke.  A more detailed description of the CMS
detector, together with a definition of the coordinate system used and
the relevant kinematic variables, can be found in
Ref.~\cite{:2008zzk}.

The CMS experiment uses a two-level trigger system. The level-1 (L1)
trigger~\cite{L1TDR}, composed of custom hardware processors, selects
events of interest using information from the calorimeters and muon
detectors and reduces the readout rate from the 20\unit{MHz}
bunch-crossing frequency to a maximum of 100\unit{kHz}.  The high
level trigger (HLT)~\cite{HLTTDR} uses software algorithms accessing
the full event information, including that from the inner tracker, to
reduce further the event rate to the 400 Hz that is recorded.

\section{Methodology}

The searches described in this paper probe fundamentally different
manifestations of physics beyond the standard model. The first type
looks for a resonance structure appearing above a smooth
background. This search parametrizes the expected signal and
background shapes using appropriate functional forms and takes an
unbinned likelihood approach to establish compatibility with the
representative new physics models used as benchmarks.  The second type
looks for a smooth deviation from the background where no resonance
structure is expected. In the non-resonant analyses the number of
events above a particular invariant mass is compared with the total
number of expected background events.  The same observed mass spectra
are used by all of the analyses.

To be robust against uncertainties in the absolute background level,
the search for resonances makes use only of the shape of the dilepton
mass spectra. In the absence of a signal, limits are set on the ratio
$R_{\sigma}$ of the production cross section times branching fraction
for high-mass resonances to that of the $\cPZ$ boson.  In this
approach, many experimental and theoretical uncertainties common to
both measurements cancel or are reduced. The non-resonant analyses
also use the number of reconstructed $\cPZ$ boson events to reduce
some systematic uncertainties.  Using theoretical cross sections,
lower mass limits, or limits on model parameters in the case of the
non-resonant analyses, are calculated for specific models.  The
experimental limits derived within the resonance analysis are designed
to be easily reinterpretable in the context of any model predicting a
narrow resonance, and spin-specific parametrizations for the product
of the acceptance and the reconstruction efficiencies are provided for
completeness.
Similarly, the signal cross section limits above different lower mass
thresholds may be reinterpreted in the context of other models
predicting a non-resonant enhancement at large masses in the dilepton
mass spectrum.

\section{Event selection \label{sec:lepton}}

\subsection{Triggers \label{sec:triggers}}

The trigger used to select dimuon events requires at least one muon
candidate with transverse momentum $\pt > 40$\GeV.  The candidate muon
tracks in the HLT are created by combining tracks reconstructed using
muon chamber information alone with information from the silicon
tracking detectors.  To keep the trigger rate at an acceptable level,
the acceptance of this trigger is restricted to a pseudorapidity range
of $\abs{\eta} < 2.1$. In addition, the candidate tracks are required to
have a $\chi^2/\mathrm{dof} < 20$ and to have a point of closest
approach to the beam axis of less than 0.1\unit{cm} in the transverse
plane.

The trigger used to select dielectron events requires the presence of
two clusters in the ECAL, each associated with a track reconstructed
using tracker information. The clusters are reconstructed by summing
energy deposits in crystals surrounding a ``seed'', which is locally
the crystal containing the largest energy. The summing procedure
encompasses energy deposits potentially arising from bremsstrahlung
emission. The clusters are required to have transverse energies \ET
($=E\sin(\theta)$) greater than $33\GeV$. The total energy in the
hadron calorimeter cells, within a cone of radius $\Delta R =
\sqrt{\smash[b]{(\Delta\eta)^2 + (\Delta\phi)^2}} < 0.14$ centred on the ECAL
cluster, is required to be less than 15\% (10\%) of the cluster energy
in the barrel (endcap) region of the ECAL. At least one of the ECAL
clusters identified in the HLT is required to be compatible with an
energy deposit identified by the L1 trigger. In the electron trigger,
ECAL cluster information is used to identify associated hits in the
pixel detector, which are then used to initiate track reconstruction.

The trigger efficiencies can be represented as a product of
uncorrelated efficiencies of the different trigger components. In order to
determine these separate elements, various triggers with criteria
different from those used for the signal triggers are employed. In
general, these triggers have larger rates than the signal triggers and
are prescaled, meaning that only a fraction of the events potentially
passing these triggers are recorded. The tag-and-probe methodology
described in Refs.~\cite{EWK-10-002-PAS,MUO-10-004-PAS,CMS-dilep-2012}
is used, where applicable, to obtain detailed efficiencies for the
main contributions to the total efficiency.

The efficiency of a single muon trigger varies as a function of
$\eta$, resulting in an efficiency for triggering on a dimuon system
that varies between 97 and 100\%.  From simulations this efficiency is
constant over the mass range from the region of the $\cPZ$ peak to
greater than 3\TeV.

The total electron trigger efficiency, for events with two electron
candidates that pass the offline electron selection requirements, is
$(99.3\pm0.1)\%$ for $\ET > 38\GeV$, where the trigger efficiency
reaches a plateau.  Where required in the dielectron resonance
analysis, and in the plots relevant to electrons shown in this paper
comparing predicted event yields with data, the effect of trigger
efficiencies on simulated event samples is included by using the
trigger efficiencies determined from the data and applying a weight to
each simulated event. In the ADD and CI analyses a systematic
uncertainty is assigned to account for the small inefficiency.

\subsection{Lepton reconstruction and identification \label{sec:leptonID}}

Muons and electrons are reconstructed using standard algorithms,
described in more detail in
Refs.~\cite{MUO-10-004-PAS,EWK-10-002-PAS,CMS-dilep-2012}.  The
primary vertices in the event are reconstructed using silicon tracker
information ~\cite{CMS-trk-paper}.  In each analysis, the primary
vertex closest to the origin of the reconstructed pair of leptons is
used.

Muon tracks are reconstructed separately in both the muon system and
the silicon tracker. For each compatible pair of tracks the set of
space points is fitted to form a track that spans the entire
detector~\cite{MUO-10-004-PAS}.
For muons with $\pt<200\GeV$, the transverse momentum
resolution is dominated by the resolution on track parameters in the
inner tracker~\cite{MUO-10-004-PAS}. However, above 200\GeV the muon
stations also contribute significantly to the precision
of the measurement.  The muon \pt
resolution is estimated using collision data and cosmic rays to be
around 2\% for $\pt\approx100\GeV$ and better than 10\% for
$\pt\approx1\TeV$, for muons reconstructed in the barrel. Monte Carlo
simulations reproduce the performance observed in collision data and
cosmic rays.
Each of the muon candidates is
required to have $\pt > 45\GeV$ and $\delta(\pt)/\pt < 0.3$, where
$\delta(\pt)$ is the uncertainty in the measured \pt of the track.
The muons must lie within the acceptance of the muon detectors,
$\abs{\eta}<2.4$, furthermore the muon that triggers the event must be
within $\abs{\eta}<2.1$ as a consequence of the trigger criteria. The muon
candidates are required to have a transverse impact parameter of less
than 0.2\unit{cm} with respect to the primary vertex position, at
least one hit in the pixel detector, hits in at least six
silicon-strip tracker layers, and matched segments in two or more muon
stations.  To suppress backgrounds from non-prompt muons, the scalar
sum of the \pt of all other tracks with a $z$ impact parameter within
0.2\unit{cm} of the relevant primary vertex and lying within a cone of
$\Delta{R} < 0.3$ about the track of the muon candidate, is required
to be less than 10\% of the \pt of the candidate.
The impact parameter criterion also reduces the effect of tracks
originating from additional $\Pp\Pp$ interactions occurring in the
same bunch crossing (pileup) on reconstructed quantities.  Its
effectiveness was assessed using muons arising from Z bosons where
efficiencies have been shown to agree in data and simulation. When
varying the average number of pileup events between 0 and 30 a change
of less than 1\% is observed in the muon selection efficiency.

Clusters in the ECAL are matched to hits in the silicon pixel
detector, which are then used to seed tracks in the rest of the
tracker.  The resulting cluster-track matched pairs form electron
candidates.  These candidates are required to have $\ET >35\GeV$ and
$\abs{\eta}<2.5$, excluding the barrel-endcap transition region
$1.442<\abs{\eta}<1.560$. To suppress the misidentification of jets as
electrons, the sum of the $\pt$ of all other tracks in a cone of
$\Delta R < 0.3$ around the track of the electron candidate is
required to be less than 5\GeV, which imposes an isolation condition
on the track.  To be used in the calculation of the isolation of the
candidate track, the tracks have to be within 0.2\cm, in the $z$
direction, of the primary vertex with which the electron candidates
are associated. This requirement reduces the impact pileup.
For electrons with transverse energies above 100 GeV, a
negligible change in the selection efficiency is observed as the
number of pileup events increases from 0 to 40.  For electrons
identified as arising from Z bosons, i.e. where the $\ET$ are much
lower than 100 GeV, the efficiency falls by between 5 and 10\%
depending on the region of the detector in which the electrons are
detected.
Within this same cone, the sum of the
$\ET$ of the energy deposits in the calorimeter that are not
associated with the candidate is required to be less than 3\% (plus a
small $\eta$-dependent offset) of the candidate $\ET$. This sum, which
allows a selection on the isolation of the electron candidate, is
corrected for the average energy density in the
event~\cite{Cacciari:2007fd} to minimize the dependence of the
efficiency of this selection criterion on pileup.
Further suppression of the misidentification of jets as electrons is
achieved by requiring that the profile of the energy deposition in the
ECAL be consistent with that expected for an electron, and that the
sum of HCAL energy deposits in a cone of $\Delta R<0.15$ be less than
5\% of the ECAL energy of the electron. The track associated with the
cluster is required to have no more than one hit missing in the pixel
layers, and in the transverse plane to lie within 0.02~cm (barrel) or
0.05\unit{cm} (endcaps) of the primary vertex associated with the
candidates.
The energy resolution for the selected electrons varies between approximately
1.0 and 3.5\% depending on the momentum, the extent of bremsstrahlung
emission and the point
of incidence on the ECAL~\cite{CMS-elec-paper}.

For signal events, the total efficiency (including triggering,
reconstruction, and identification) is estimated from simulated
events. In the resonance analysis, limits are set on $R_{\sigma}$,
which is the ratio of the product of cross section and branching
fraction for $\zp$ production relative to that for $\cPZ$ bosons.  In
the non-resonant analyses the relevant ratio is the one between the
summed events above the minimum mass threshold and the number of
events in the $\cPZ$ peak region.  Therefore the simulation does not
need to reproduce the absolute value of the efficiency in data; it
must however correctly reproduce the evolution of the efficiency with
\ET. Data are used to measure the electron identification efficiency
at the \Z resonance, using the tag-and-probe
method~\cite{EWK-10-002-PAS,MUO-10-004-PAS,CMS-dilep-2012}.  The ratio
of this efficiency to that found in simulation is $0.997\pm0.007$ for
electrons in the barrel region and $0.979\pm0.006$ in the endcaps.
This ratio was also studied as a function of the probe electron \ET to
$\sim$500\GeV and as a function of the tag-and-probe pair mass up to
$\sim$1\TeV, and was found to be invariant with respect to these
quantities, although the systematic and statistical uncertainties
become large at high transverse energies and masses.  Using these
scale factors, the total efficiency to reconstruct and select
electrons with $\pt > 100\GeV$ is expected to be $(88\pm2)\%$ in the
barrel region and $(84\pm4)\%$ in the endcaps,
where the uncertainties cover the extrapolation of the data-to-simulation ratio to very high transverse momenta.
This gives efficiencies
for electron pairs of 78\% where both electrons are in the barrel and
75\% where one of the electrons is in the endcap.  A similar procedure
is used to evaluate the muon identification efficiency.  Applying the
tag-and-probe technique to muons from $\Z$ boson decays, and using
tracks in the silicon inner tracker as probes, the total muon
identification efficiency (including isolation) is measured to be
$(95\pm1)\%$ in the barrel and endcap regions.  The corresponding
efficiency ratios between data and the simulation are $0.990\pm0.005$
and $0.993\pm0.005$, respectively.  To within the statistical
precision available using the 2012 data sample, both the efficiencies
and the related correction factors remain constant up to a \pt of
approximately 300\GeV.  With these correction factors applied, the
combined reconstruction and selection efficiency for triggered events
in the acceptance region is expected to be $(89\pm2)\%$ at a mass of
200\GeV. The simulation predicts that the efficiency above 200\GeV is
constant to within 3\%.

For both the dimuon and dielectron final states, two isolated,
same-flavour leptons that pass the lepton identification criteria
described above, are required. In the case of muons, the particle
momentum is measured from the curvature of the associated track, and
thus the sign of the charge is automatically determined. Muons in
dimuon events are therefore required to have charges of opposite sign.
Dielectron events are separated into barrel-barrel and barrel-endcap
categories, because of the different signal-to-background ratios and
mass resolutions in the two regions.  The measurement of the energy of
the electron depends only upon the energy deposition in the ECAL and
therefore is independent of the charge of the particle. The
opposite-sign requirement is therefore not applied to dielectron
events since this would only result in a loss of efficiency without
providing useful background suppression.

Muon candidates are also required to originate from the same vertex,
by requiring the $\chi^2/\mathrm{dof}$ to be less than 10 when tracks
are fitted to a common vertex.  Cosmic ray and beam halo muons are
suppressed to negligible levels by the common-vertex requirement
together with the further requirement that the opening angle between
the two reconstructed muons be less than $\pi - 0.02$ radians.

The acceptance times efficiency for heavy particles with spins 1 and 2
decaying to lepton pairs is determined using simulation and varies as a
function of the particle mass. Table~\ref{tab:accEff} lists the
parametrizations of the acceptance times efficiency as a function of
mass along with their evaluation at two example mass points, 1000 and
2500\GeV. The uncertainties in this table are dominated by the
uncertainties in the efficiencies described above.

\begin{table*}[htb!]
\centering
\caption{Parametrizations of the product of the acceptance and the
  efficiency as a function of the mass $m$ expressed in units of \GeVns.}
  \renewcommand{\arraystretch}{1.3}
\begin{tabular}{c|cccc}
\multicolumn{1}{c}{} & \multirow{2}{*}{Functional form} & \multicolumn{2}{c}{Acceptance$\times$Efficiency} &\multirow{2}{*}{Uncertainty} \\
\cline{3-4}
 \multicolumn{1}{c}{}&\multicolumn{1}{c}{} & \multicolumn{1}{c}{1000\GeV} & \multicolumn{1}{c}{2500\GeV} & \multicolumn{1}{c}{}\\\hline
spin 1   & & & & \\
\hline
$\Pgmp\Pgmm$                 & $0.81 - \frac{1.5\times 10^8}{ (m+570)^3}  $                & 0.77 & 0.80 & 3\% \\
ee barrel-barrel                 & $0.59 - \frac{2.9 \times 10^5}{m^2 + 7.5 \times 10^5} $            & 0.43 & 0.55 & 4\%  \\
ee barrel-endcap                 & $ 0.06 - \frac{159}{m+345} + \frac{7.3\times 10^5}{m^2 + 1.8 \times 10^6}$     & 0.21 & 0.10 & 6\% \\
\hline\hline
spin 2   & & & & \\
\hline
$\Pgmp\Pgmm$                 & $0.75 + \frac{112}{m -  104} - \frac{6.1 \times 10^4}{m^2 + 1.6 \times 10^4}$ & 0.82 & 0.79 & 3\% \\
ee barrel-barrel                 & $0.57 - \frac{3.0 \times 10^4}{m^2 + 1.3 \times 10^5}$                        & 0.54 & 0.57 & 4\%\\
ee barrel-endcap                 & $-0.24 + \frac{1.2 \times 10^4}{m + 3.5 \times 10^4} + \frac{6.6 \times 10^4}{m^2 + 7.5 \times 10^5}$ & 0.13 & 0.09& 6\% \\
\end{tabular}
\label{tab:accEff}
\end{table*}

\subsection{Study of energy deposits in the ECAL crystals}

The energy calibration procedure for the crystals in the ECAL makes
use of both collision data and test beam data taken before the
installation of the detector~\cite{:2008zzk}. This procedure provides
a well-calibrated detector up to energies of around 100\GeV. Very high
energy electrons can potentially deposit energies of several hundreds
of \GeVns{} in a single crystal. The procedure described below was
developed in order to ensure that the calibration could be extrapolated
to such energies.

An electron deposits its energy over an array of crystals
approximately 5 $\times$ 5 in extent. In general, only the energy
deposited in the central crystal will significantly exceed the
energies at which calibrations are performed.  Given the shape of the
electromagnetic shower, the energy of the central crystal can be
predicted from the energy distribution in the surrounding 24 crystals.
This is done using both an algorithmic fit to the distributions and an
artificial neural network.  The difference between the observed energy
$E_1$ in the central crystal and the energy reconstructed,
$E_1^{\text{rec}}$, is calculated from data using the above method.
The distribution $(E_1-E_1^{\text{rec}})/E_1$ obtained using
simulated data results in a displacement from zero of the mean of the
distribution of less than 0.01. For electrons in the data with
energies above 500\GeV, the distribution found using simulation agrees
with that for the data, where displacements of the mean are observed
to be less than 0.01 in both the barrel and endcap regions. This procedure
has also been adapted to reconstruct the energies in any crystals in a
cluster where the readout is not functioning. This correction is used
in about 2\% of events where the dilepton mass is above 200\GeV.

\section{Background sources}

The principal standard model process that contributes to the dimuon
and dielectron invariant mass spectra, either directly or via
$\tau\tau$, is Drell--Yan production ($\cPZg$). There are also
contributions from $\ttbar$, $\cPqt\PW$, and diboson processes. In
addition, jets may be misidentified as leptons and contribute to the
dilepton invariant mass spectra through multijet and vector boson plus
jets final states. The contribution from diphotons misidentified as
dielectrons
 has been established to be negligible.

 \subsection{The \texorpdfstring{\Z/$\gamma^*$}{Z/gamma*} background}

Drell--Yan production is simulated using the \POWHEG
(r1513) \cite{Nason:2004rx,Frixione:2007vw,Alioli:2010xd}
next--to--leading order (NLO) event generator.
 The differential cross section from this source can be modified by
higher-order corrections and by variations in Parton Distribution
Functions (PDF). Uncertainties due to these sources are included in
all of the analyses and are described in
Section~\ref{sec:uncertanties}. The resonance search relies only on
the shape of the spectrum, using data primarily from mass regions
where no non-standard-model contribution is expected, to constrain the
magnitude of the background, making the result insensitive to the
predicted absolute cross section.  The non-resonant analyses use
simulation to predict the number of events and hence are more
sensitive to these systematic uncertainties.

\subsection{Other background sources with prompt lepton pairs
\label{sec:e-mu}  }
Prompt lepton pairs can result from {$\ttbar$}, $\cPqt\PW$, and diboson
production in addition to the dominant Drell--Yan process. In order to
demonstrate that simulations provide a good representation of these
processes, their flavour-symmetric nature is exploited. The
$\Pe^{\pm}\Pgm^{\mp}$ invariant mass spectrum is used solely to test
the quality of the simulations.  The {$\ttbar$} and $\cPqt\PW$ simulated
samples are generated using \POWHEG~\cite{Re:2010bp}, while the
diboson simulated samples are generated using the \PYTHIA
v6.426~\cite{Sjostrand:2006za} event generator.

\begin{figure}
\centering
 \includegraphics[width=\cmsFigWidth]{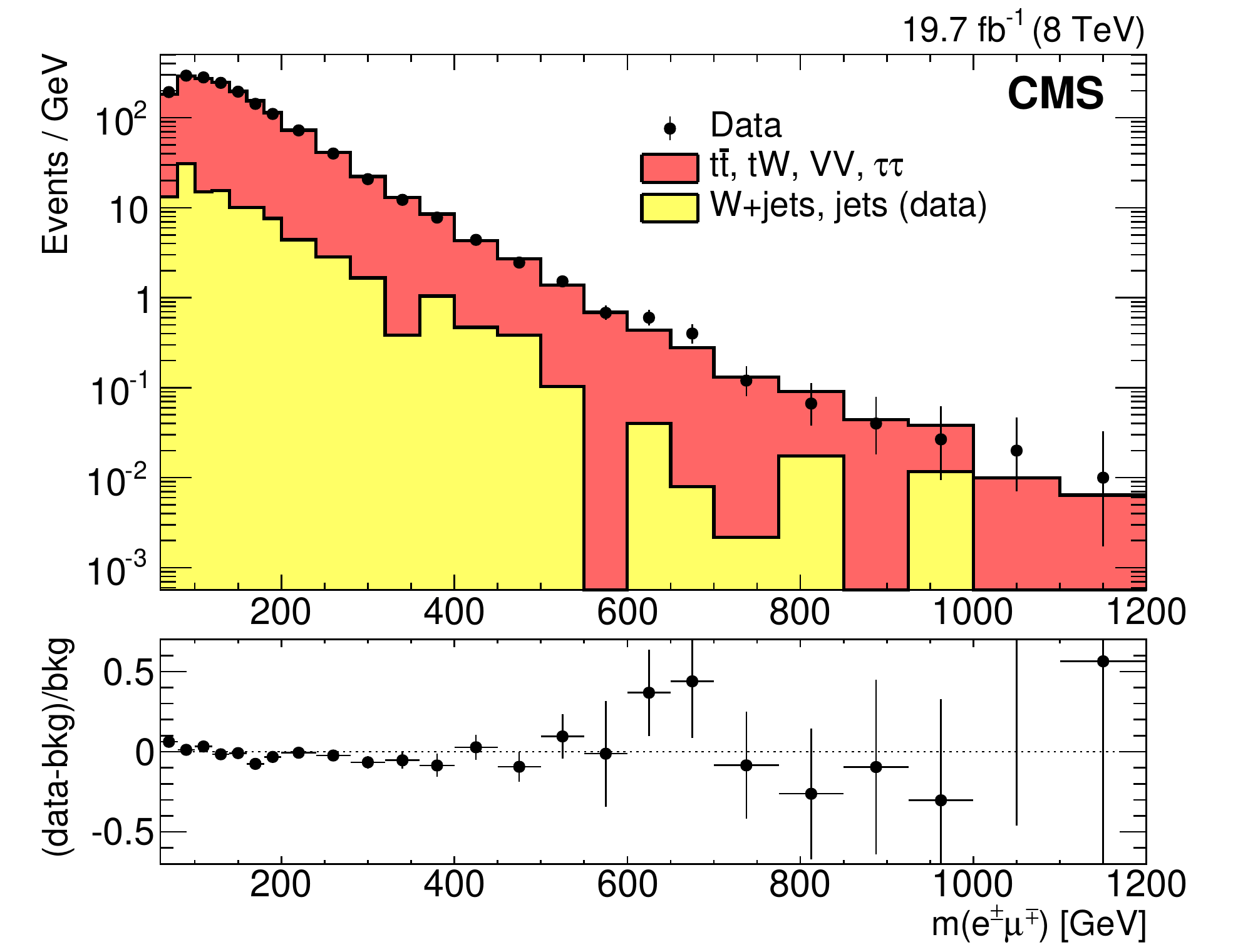}
\caption{\label{fig:muonselectrons} The observed opposite-sign
  $\Pe^\pm\Pgm^\mp$ dilepton invariant mass spectrum (data points).
  The filled red (dark shading) histogram shows the contribution to
  the spectrum from $\ttbar$ and other sources of prompt leptons
  ($\cPqt\PW$, diboson production, $\ztt$, $\PW$+jets, jets).  The
  background where at least one of the reconstructed objects is not a
  correctly identified lepton is shown in yellow (light shading). All
  components are estimated using simulations except for the jet
  component where both leptons are misidentified jets, which is
  estimated from the data using the same-sign $\Pe^\pm\Pgm^\pm$
  spectrum. The simulated backgrounds are normalized so that in the
  dielectron channel, the observed data and the prediction from
  simulation agree in the region of the $60<m_{\Pe\Pe}<120\GeV$.
  The lower plot shows the difference between the number of data and
  background events in each bin divided by the number of background
  events. All error bars shown are statistical only. }
\end{figure}

Figure~\ref{fig:muonselectrons} shows the $\Pe^{\pm}\Pgm^{\mp}$
invariant mass spectrum resulting from a trigger that requires the
presence of both a muon and an electromagnetic object. The muon and
electron selection criteria described in Section~\ref{sec:lepton} are
used and the leptons are required to have opposite signs.  All
components are estimated from simulations except for the component
arising from multijet events where both lepton candidates are
misidentified jets.  This component is derived from data by using the
same-sign $e\mu$ spectrum. The observed number of opposite-sign
$\Pe\Pgm$ events is 20\,513 (6756) in the mass region above 120 (200)\GeV. Using simulations, and the estimation from data for the
contribution from jets being misreconstructed as leptons, the
expected number of events above 120 (200)\GeV is $21\,100\pm600$
($6800\pm200$).

\subsection{Events with misidentified and non-prompt leptons}

Both jets and photons can be misidentified as prompt
electrons. Potential sources of such backgrounds are
$\PW(\to\Pe\cPgn)$+jets, $\gamma$+jets, and multijet events. The
method described below primarily uses the data to determine the
contribution to the observed mass spectra from these sources.  The
misidentification rate (MR) is the probability for a jet, having been
reconstructed as an electron candidate, to pass the electron
selection.  This rate is measured using a sample collected with a
prescaled single electromagnetic cluster trigger. To suppress the
contribution from $\cPZ$ boson decays, events in this sample are
required to have no more than one reconstructed electron above 10\GeV.
Contamination from genuine electrons in $\PW$+jet events and from
converted photons in $\gamma$+jet events may affect the MR
measurement. A less significant source of contamination is from
processes that can give a single electron such as \ttbar, tW, WW, WZ,
$\ztt$, and $\zee$, and in which a second electron is produced but fails
to be reconstructed.  The effect of the real electromagnetic object
contamination on the MR is corrected using simulated samples where the
$\PW$+jet simulated sample is generated using
\MADGRAPH~5~\cite{MadGraph5} and the $\gamma$+jet sample is derived
from the \PYTHIA event generator. Following these corrections, the MR
is defined as the number of electrons passing the full selection
divided by the number of electron candidates in the sample. The
misidentification rate is quantified in bins of \ET and $\eta$.

Once this rate has been measured, the jet background can be estimated
using two samples selected from data. The first sample consists of
events with two reconstructed electron candidates that pass the
trigger, but fail the full selection. The second sample is similar to
the first sample except one electron has to pass the full selection
instead of failing it. When weighted by the $\mathrm{MR}/(1-\mathrm{MR})$
appropriate for each electron, the first sample estimates the multijet
component, where both electrons candidates arise from misidentified
jets, of the jet background only. When weighted by the $\mathrm{MR}/(1-\mathrm{MR})$
appropriate for the failing electron, the second sample estimates the
sum of the $\PW\to \Pe\cPgn$+jet background, $\gamma$+jet
background, and twice the multijet background. The second sample
overestimates the multijet background by a factor of two because there
are two combinations possible with one electron passing and one
failing the selection. However, as the multijet background has been
estimated by the first sample, the total jet background is the number
of events estimated from the second sample minus the number of events
estimated from the first sample. Based on the results of various
consistency tests and cross-checks, a 40\% systematic uncertainty is
applied to the misidentified-jet background estimate. The estimated
background contributions are shown in Table~\ref{tab:event_yieldee}.

In principle the dimuon channel also has contributions from jets and
photons that are misidentified as muons. However, the background from
this source, determined using the procedure in
Ref.~\cite{CMS-dilep-2012}, is found to be negligible as can be seen in
Table~\ref{tab:event_yieldmumu}.

\subsection{Cosmic ray muon backgrounds}

The potential background in the $\Pgmp\Pgmm$ sample from events
containing cosmic ray muons is suppressed by the selection criteria
described in Section~4.2, which reject events with back-to-back muons
and events with muons that have a large impact parameter relative to
the collision vertex.  For the dimuon mass region $m_{\Pgm\Pgm} >
200$\GeV, the residual expected background is estimated using two
event samples.  Events in one sample are selected without imposing the
requirement on the dimuon opening angle and in the other sample the
requirements on muon impact parameter and on the existence of a good
quality primary vertex are not applied.  The efficiencies of the
remaining selection requirements are estimated using these samples
under the assumption that they are uncorrelated. The background due to
cosmic rays is estimated to be less than 0.1 events above a mass of
200\GeV.

\section{Dilepton invariant mass spectra}

\begin{figure}[htbp]
\centering
\includegraphics[width=\cmsFigWidth]{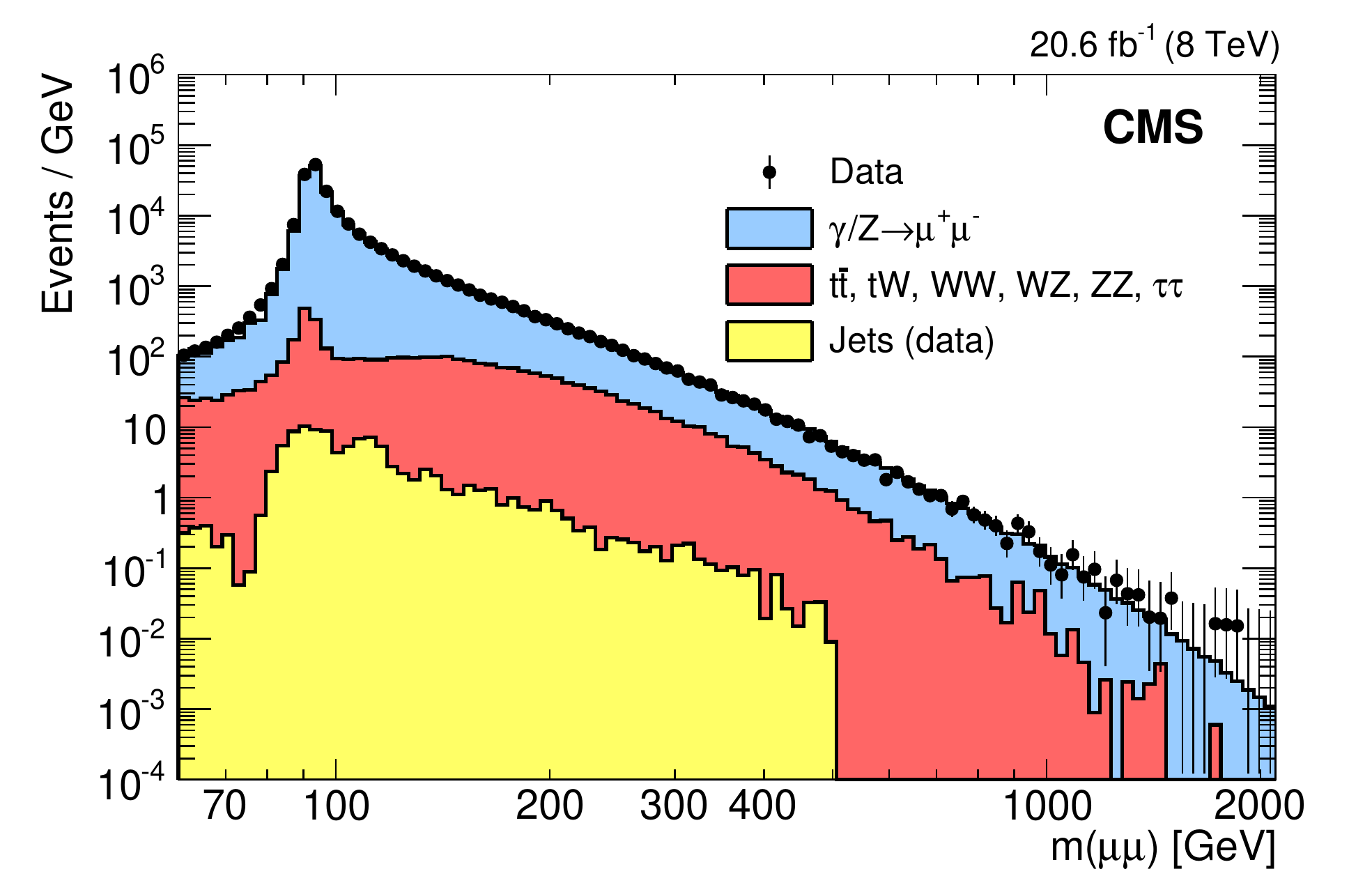}
\includegraphics[width=\cmsFigWidth]{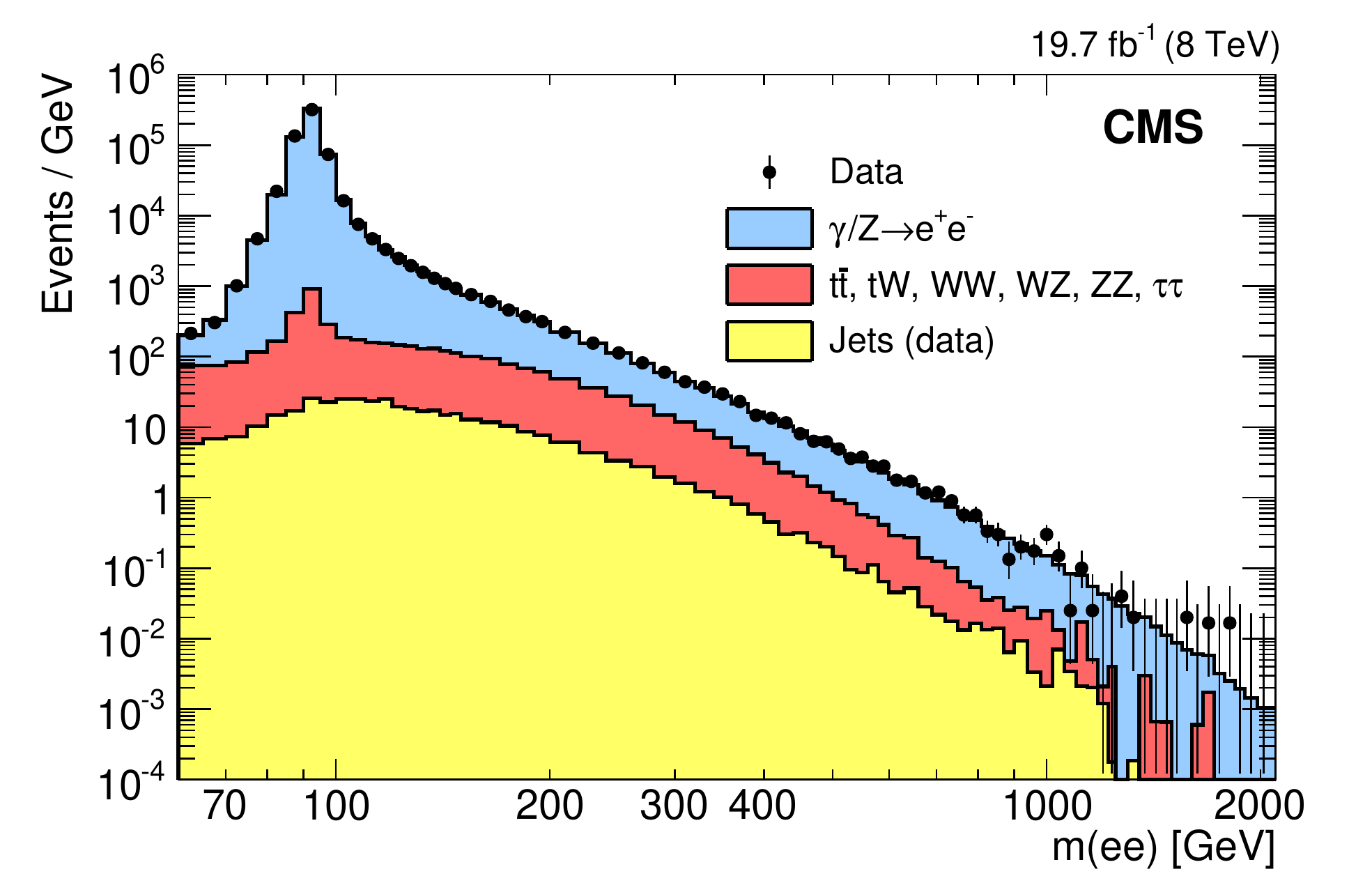}
\caption{\label{fig:spectra}
The invariant mass spectrum of $\Pgmp\Pgmm$ (top) and $\Pe\Pe$ (bottom) events. The points with
error bars represent the data.
 The histograms represent the expectations from SM processes: $\cPZg$,
$\ttbar$, and other sources of prompt leptons ($\cPqt\PW$, dibosons,
$\ztt$), as well as the multijet backgrounds. Multijet backgrounds
contain at least one jet that has been misreconstructed as a lepton.
The simulated backgrounds are normalized to the number of events in
the data in the region of $60<m_{\ell\ell}<120\GeV$, with the dimuon
channel using events collected with a prescaled lower-threshold
trigger.
}
\end{figure}

\begin{figure}[htbp]
\centering
 \includegraphics[width=\cmsFigWidth]{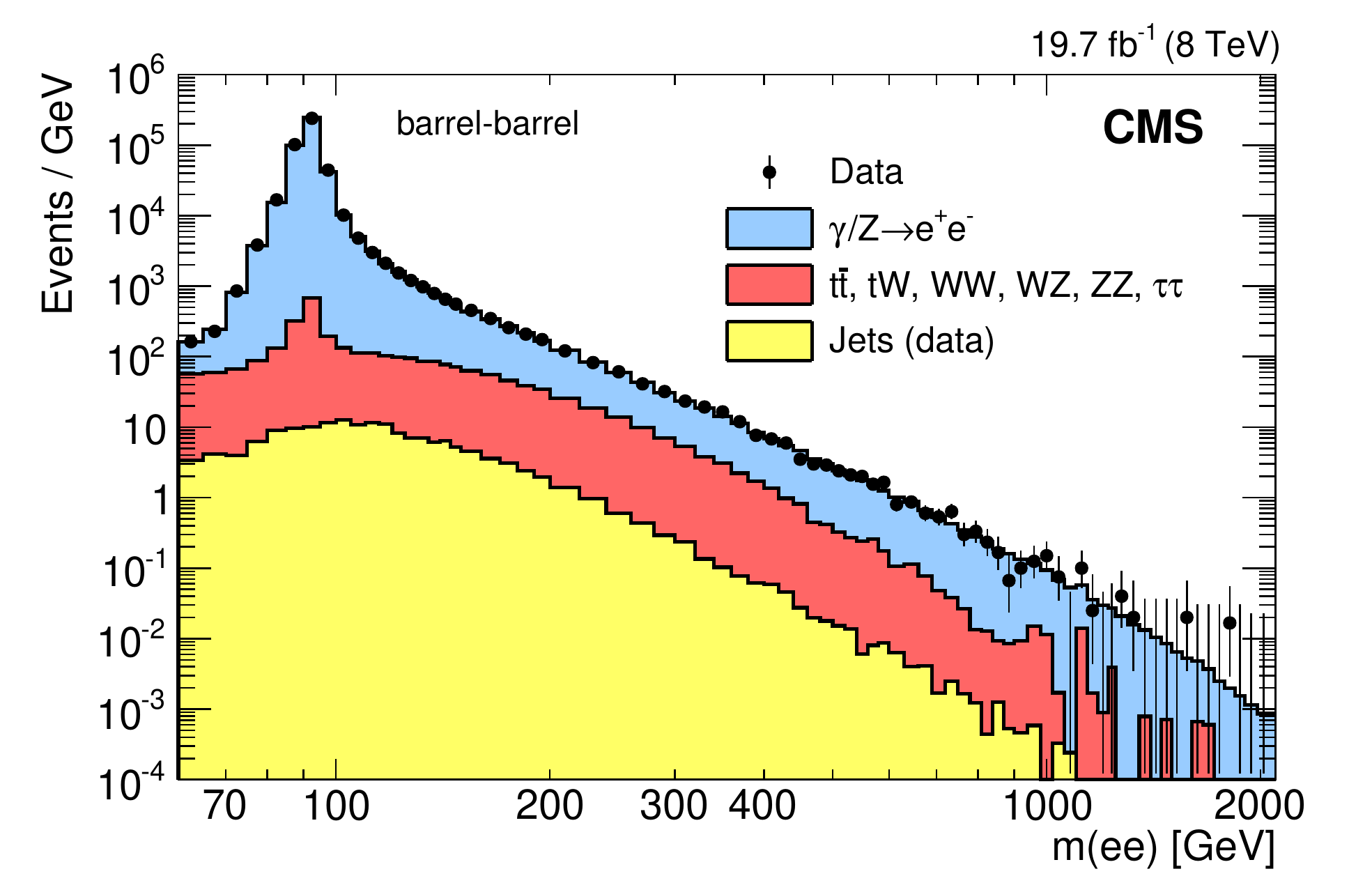}
 \includegraphics[width=\cmsFigWidth]{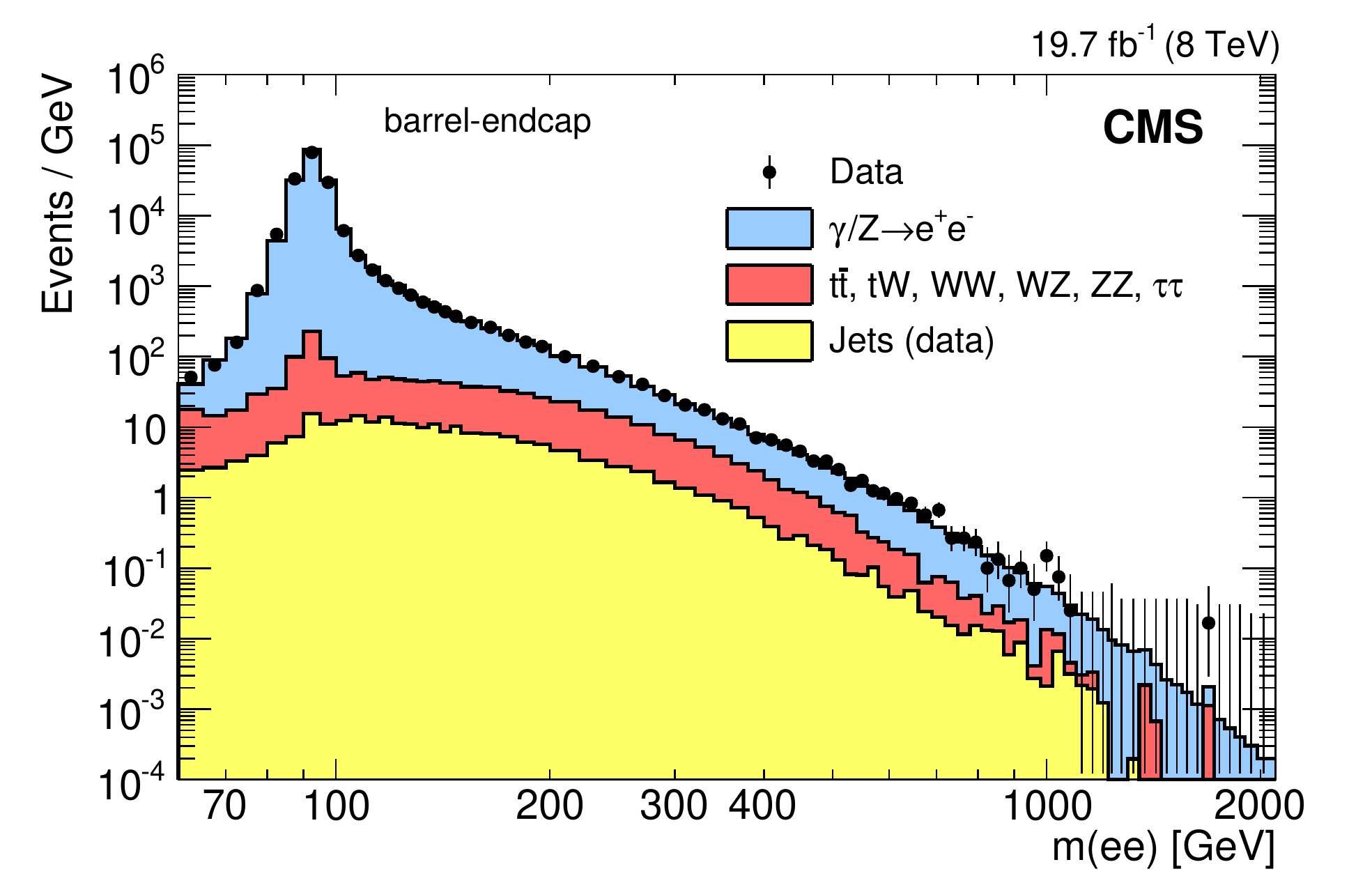}
\caption{\label{fig:spectraEleOnly}
The invariant mass spectrum for $\Pe\Pe$ events separated into
barrel-barrel (top) and barrel-endcap (bottom) categories. The points
with error bars represent the data.
The histograms represent the expectations from SM processes: $\cPZg$,
$\ttbar$, and other sources of prompt leptons ($\cPqt\PW$, diboson
production, $\ztt$), as well as the multijet backgrounds. Multijet
backgrounds contain at least one jet that has been misreconstructed
as an electron. The simulated backgrounds are normalized to the data
using events in the region of $60<m_{\Pe\Pe}<120$\GeV.
}
\end{figure}

\begin{figure}[htbp]
\centering
\includegraphics[width=\cmsFigWidth]{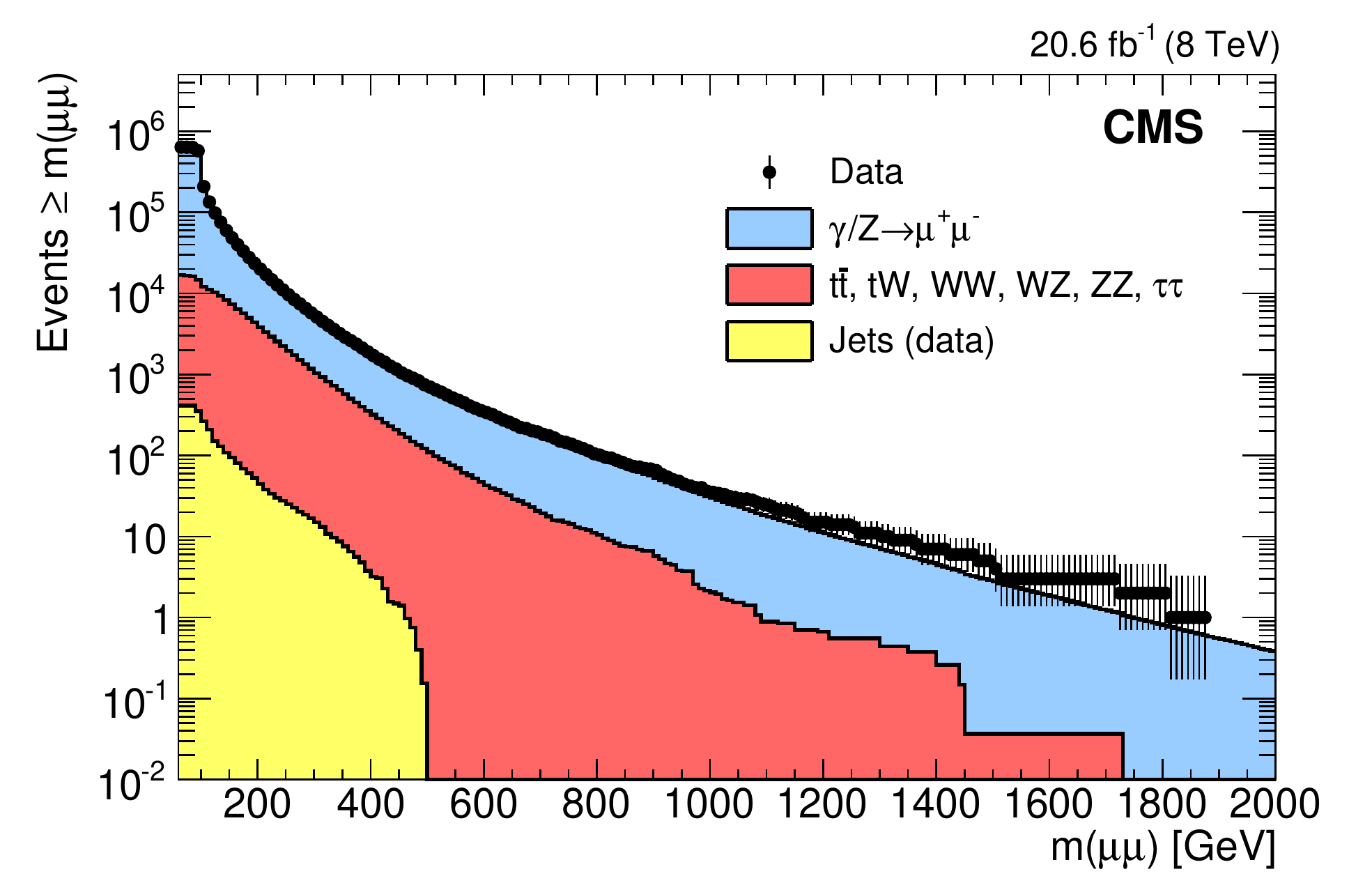}
\includegraphics[width=\cmsFigWidth]{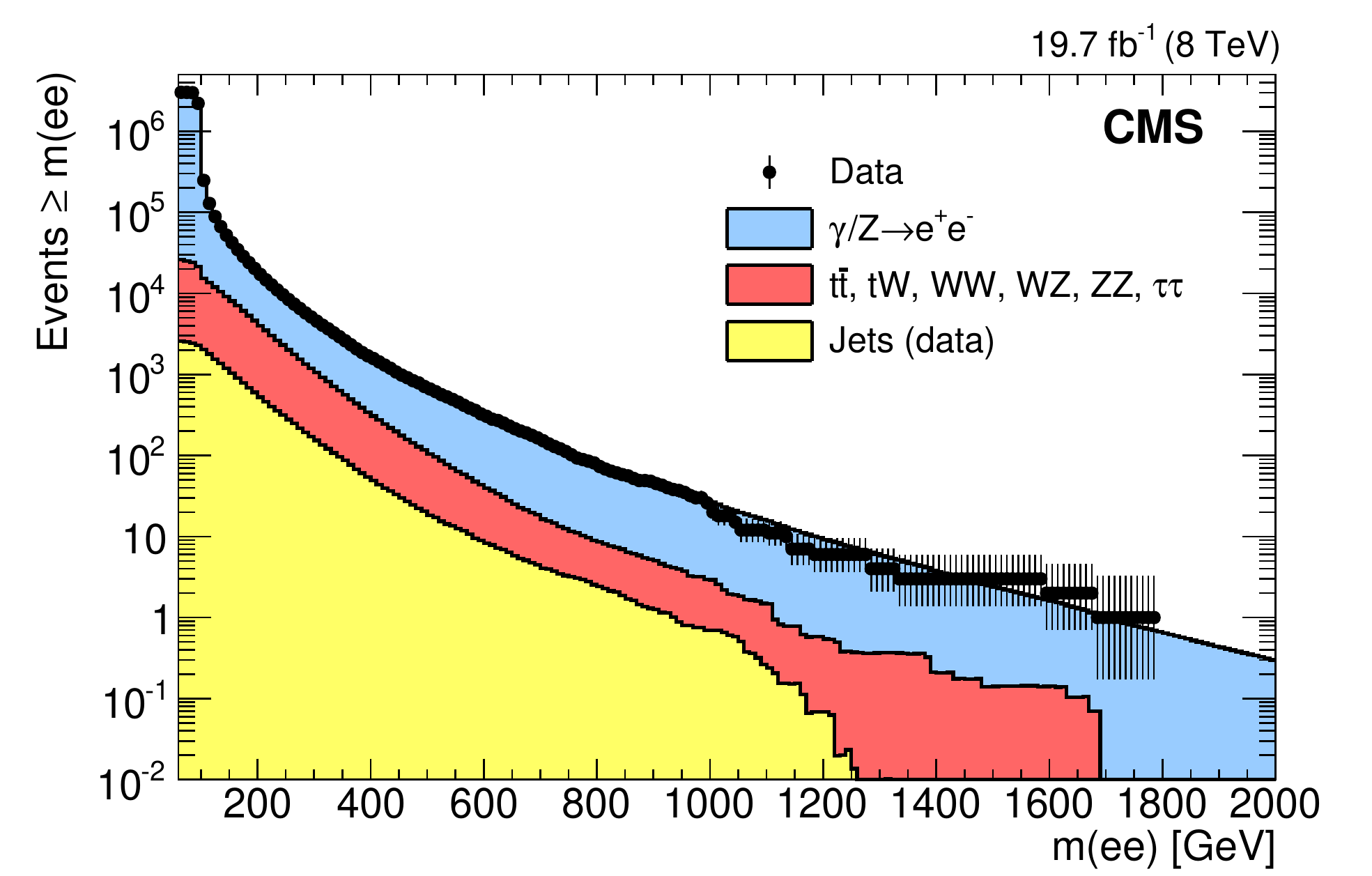}
\caption{\label{fig:cum_spectra} The cumulative distribution, where
all events above the specified mass on the $x$ axis are summed, of the
invariant mass spectrum of $\Pgmp\Pgmm$ (top) and $\Pe\Pe$ (bottom)
events.  The points with error bars represent the data; the histograms
represent the expectations from standard model processes. The
simulated backgrounds are normalized to the number of events in the data in the region of
$60<m_{\ell\ell}<120\GeV$, with the dimuon channel using events
collected with a prescaled lower-threshold trigger.}
\end{figure}

The dilepton invariant mass spectra are shown in
Figs.~\ref{fig:spectra} and \ref{fig:spectraEleOnly}, where the data
are compared with the expected backgrounds. If more than two leptons
passing all selection criteria are present in an event, the pair with
the largest invariant mass is chosen.  The largest dimuon invariant
mass observed is 1840\GeV and the largest dielectron mass is 1790\GeV.
In Figs. 2 and 3 the contribution labeled
``Jets'' consists of events where at least one jet has been
misreconstructed as a lepton.
 The other background components are derived from simulations. The
relative fractions of the simulated processes are fixed by their
theoretical cross sections, and the total simulated background
contribution is normalized to the number of events in the data in the
region of $60<m_{\ell\ell}<120\GeV$. The $\pt>45\GeV$ selection
requirement on the muons removes most of the events in this mass
range, and also has a significant effect on the efficiency of events
above this mass region (but well below the search region of interest).
To derive the normalization for dimuon production, a prescaled
trigger, which is identical to the main muon trigger but with a lower
\pt selection criterion, is used to select events. The use of this
trigger allows the offline \pt threshold criterion to be lowered to
27\GeV.  Figure~\ref{fig:cum_spectra} shows the corresponding
cumulative distributions of the spectra. The SM expected yields in
various mass bins are shown compared to the observed yields in
Tables~\ref{tab:event_yieldmumu} and \ref{tab:event_yieldee}. These
plots and the tables illustrate that there is good agreement between
observation and expectation in the whole explored region including
dilepton masses above 1\TeV. The analyses are designed to look for
evidence of beyond the standard model physics that is expected to
become manifest at masses above about 1\TeV, nonetheless the entire
region above masses of 200\GeV is examined for such evidence.

\begin{table*}[htb!]
\centering
\topcaption{
\label{tab:event_yieldmumu}
The number of dimuon events in various invariant mass ranges for an
integrated luminosity of $\anaLumimm\fbinv$.  The total background is
the sum of the events for the standard model processes listed.  The
yields from simulation are normalized relative to the expected cross
sections, and overall the simulation is normalized to the data using
the number of events in the mass window 60--120\GeV acquired using a
prescaled low threshold trigger.  Uncertainties include both
statistical and systematic components, summed in quadrature. A dash
(---) is used to indicate negligibly small contributions.  }
\begin{tabular}{c|cc|ccc}
\multicolumn{1}{c}{$m_{\mu\mu}$ range}& Data    & \multicolumn{1}{c}{Total}             & $\cPZg$           &  $\ttbar$ + other  &  Jet mis-        \\
\multicolumn{1}{c}{(\GeVns)}                &         & \multicolumn{1}{c}{background}        &                   &  prompt bkgd       &  reconstruction  \\ \hline
120--400             &  96299  &  96800 $\pm$ 4300 & 86800 $\pm$ 3800  &   9900 $\pm$ 420   &  147 $\pm$ 18    \\
400--600             &  1367   &  1460  $\pm$ 80   & 1180 $\pm$ 60     &   276 $\pm$ 13     &  3  $\pm$ 3      \\
600--900             &  273    &  283 $\pm$ 19     &  246  $\pm$ 16    &   37  $\pm$ 4      &  ---              \\
900--1300            &  55     &  46  $\pm$ 4      &   40  $\pm$ 4     &   5  $\pm$ 1       &  ---              \\
1300--1800           &  8      &  6.1  $\pm$ 0.8   &    5.7  $\pm$ 0.8 &   0.4  $\pm$ 0.2   &  ---              \\
$>$1800             &  2      &  0.8  $\pm$ 0.2   &   0.8  $\pm$ 0.2  &   ---               &  ---              \\
\end{tabular}
\end{table*}
\begin{table*}[htbp]
\centering
\topcaption{The number of dielectron events in various invariant mass
ranges for an integrated luminosity of $\anaLumiee\fbinv$.  The total
background is the sum of the events for the standard model processes
listed.  The yields from simulation are normalized relative to the
expected cross sections, and overall the simulation is normalized to
the data using the number of events in the mass window 60--120\GeV.
Uncertainties include both statistical and systematic components,
summed in quadrature.  A dash (---) is used to indicate negligibly
small contributions.  }
\begin{tabular}{c|cc|ccc}
\multicolumn{1}{c}{$m_{\Pe\Pe}$ range} & Data & \multicolumn{1}{c}{Total}             & $\cPZg$           & $\ttbar$ + other   & Jet mis-           \\
\multicolumn{1}{c}{(\GeVns)}                &         & \multicolumn{1}{c}{background}        &                   & prompt bkgd        & reconstruction     \\ \hline
120--400             & 87117   & 88700 $\pm$ 3900  & 77100 $\pm$ 3900  & 10130 $\pm$ 680    & 1500 $\pm$ 300     \\
400--600             & 1266    & 1240 $\pm$ 100    & 970 $\pm$ 100     & 226 $\pm$ 15       & 40 $\pm$ 8         \\
600--900             & 259     & 245 $\pm$ 21      & 211 $\pm$ 21      & 27 $\pm$ 2         & 7 $\pm$ 1          \\
900--1300            & 41      & 39 $\pm$ 3        & 35 $\pm$ 3        & 3.5 $\pm$ 0.2      & 1.2 $\pm$ 0.2      \\
1300--1800           & 4       & 5.2 $\pm$ 0.5     & 4.8 $\pm$ 0.5     & 0.36 $\pm$ 0.02    & 0.005 $\pm$ 0.001  \\
$>$1800             & 0       & 0.64 $\pm$ 0.06   & 0.64 $\pm$ 0.06   & ---                 & ---                 \\
\end{tabular}
\label{tab:event_yieldee}
\end{table*}

\section{Statistical analysis and results}

The observed invariant mass spectra agree with expectations based on
standard model processes. Limits have been set on the possible
presence of the narrow heavy resonances predicted in various models,
and on the excesses of the form expected in the ADD and CI models. The
procedures used are described in the following sections.

\subsection{Resonance search}

An unbinned likelihood Bayesian approach is used to set 95\%
confidence level (CL) cross section limits on possible contributions
for narrow heavy resonances.

The parameter of interest is the ratio of the cross sections for
producing dilepton final states:
\begin{equation}
\label{eq:rsigma}
R_\sigma = \frac{\sigma(\Pp\Pp\to \cPZpr+X\to\ell\ell+X)}
                {\sigma(\Pp\Pp\to \cPZ+X  \to\ell\ell+X)}.
\end{equation}

The use of this ratio $R_\sigma$ eliminates the uncertainty in the
integrated luminosity and reduces the dependence on the experimental
acceptance, trigger, and offline efficiencies.  The ratio of
acceptances of a new boson to that of the $\cPZ$ boson are calculated
using simulations. This has been done for two possible scenarios,
corresponding to spin 1 or 2 final states.  The dimuon and dielectron
channels are treated separately.  The cross sections correspond to a
mass range of ${\pm}5\%\sqrt{s}$ ~\cite{Accomando:2013} about the $\zp$
on-shell mass and a mass range of $\pm$30\GeV about the $\cPZ$ peak. This
$\zp$ mass window is designed to ensure the results are as
model independent as possible, allowing for straightforward
reinterpretation of the limits derived in this paper in the context of
models not specifically addressed.  The window chosen reduces
model-dependent effects such as $\cPZg/\zp$ interference and
low mass tails due to higher parton luminosities at lower values of
$\sqrt{\hat{s}}$.

The statistical procedure presented in this section is identical to
that used in Refs.~\cite{CMS-dilep-2012,CMS-dilep-2010}, with the
exception that barrel-barrel and barrel-endcap electron events are now
treated as separate channels. This is because the signal-to-background
ratios and mass resolutions differ in the barrel and endcap regions,
and the mass limits are sensitive to these quantities. As an example,
at 500\GeV, the mass resolution in the barrel-barrel channel is 1.2\%
while it is 1.9\% in the barrel-endcap channel.  These resolutions
remain fairly constant above 500\GeV.  At high masses where there is
little background, the separation into the two channels does not
result in significantly improved limits.  The dimuon mass resolution
is similar in all regions and hence no differentiation into separate
regions is used.  The resolution varies with mass, ranging from
approximately 3\% at 500\GeV to 9\% at 3\TeV.
In the case of electrons, the mass resolution is dominated by the
energy resolution, and for muons by the momentum resolution.
A key feature of the limit-setting procedure is that it requires no
knowledge of the integrated luminosity as the background estimations
are normalized to the data in regions where contributions from SM
processes are large compared to any potential signal and the limits
are set on the ratio of $\zp$ to $\cPZ$ cross sections.

\subsection{Resonance search likelihood function}
\label{likelyhood}

The extended unbinned likelihood function for the invariant mass
spectrum consists of a sum of probability density functions (pdf) for
the signal and background shapes, where the signal fraction is set to
zero for the background-only hypothesis.  The signal is parametrized
as a convolution of a Breit--Wigner (BW) and a Gaussian resolution
function. The BW width is sufficiently small that the detector
resolution dominates.

The Poisson mean of the signal yield is $\mu_\mathrm{S} = R_\sigma
 \mu_\Z \cdot R_\epsilon$, where $R_\sigma$ is defined in
Eq.~(\ref{eq:rsigma}) and $R_\epsilon$ is the ratio of the selection
efficiency times detector acceptance for the $\zp$ decay relative to
that of the $\cPZ$ decay. The variable $\mu_\Z$ is the Poisson mean of
the number of $\cPZ\to\ell\ell$ events.  It is estimated by counting
the number of events in the $\cPZ$ boson mass region, where the
background contamination is predicted to be small (${\approx}0.5\%$ in
simulation). The quantities $\mu_\Z$ and $R_\epsilon$ are obtained
separately for the dimuon and dielectron channels.

A background pdf $f_\mathrm{B}$ is chosen and its shape parameters
fixed by fitting to the full SM background estimate in the mass range
$200 < m_{\ell\ell} < 3500\GeV$. The functional form used for the
background is $m^{\kappa}\re^{\alpha m + \beta m^{2}}$. This form
includes an $m^2$ term, which previous versions of this
analysis~\cite{CMS-dilep-2012} did not include.  The increasing range
of the fit has required this addition to ensure that the background is
well described across the entire range.  The parameters $\alpha$,
$\beta$, and $\kappa$, obtained from fitting to the simulation and
subsequently used in the limiting setting procedure, are shown in
Table~\ref{tab:backparams}.  The simulated background distributions
used to obtain the expected limits are normalized to data in the mass
regions above 200\GeV in each channel.
\begin{table}[h!]
\centering
\topcaption{The parameters $\alpha$, $\beta$ and $\kappa$  in the
  background function $m^{\kappa}\re^{\alpha m + \beta m^{2}}$ obtained from a fit to the
background simulation, where the mass $m$ is in \GeVns{}.
}
\begin{tabular}{cccc}
  \multicolumn{1}{c}{}   &  $\alpha$ ($\GeVns^{-1}$)   & $\beta$  ($\GeVns^{-2}$)   & $\kappa$ \\
\cline{2-4}
 $\Pgmp\Pgmm$                &  $-2.29 \times 10^{-3}$  & $3.32 \times 10^{-8}$   &  $-3.65$ \\
ee barrel-barrel         &  $-1.16 \times 10^{-3}$   & $-2.02 \times 10^{-7}$  & $-3.97$  \\
ee barrel-endcap         &  $-3.79 \times 10^{-3}$   & $ 1.86 \times 10^{-7}$   &  $-3.15$ \\
\end{tabular}
\label{tab:backparams}
\end{table}

Finally, the extended likelihood is:
\begin{equation}
\label{eq:likelihood}
\mathcal{L}({\boldsymbol m}|R_\sigma,M,\Gamma,w,\alpha,\beta,\kappa,\mu_\mathtt{B}) =
\frac{\mu^N \re^{-\mu}}{N!}\prod_{i=1}^{N}\left(
\frac{\mu_\mathtt{S}(R_\sigma)}{\mu}f_\mathtt{S}(m_i|M,\Gamma,w)+
\frac{\mu_\mathtt{B}}{\mu}f_\mathtt{B}(m_i|\alpha,\beta,\kappa)
\right),
\end{equation}
where ${\boldsymbol m}$ denotes the data set in which the observables
are the invariant mass values of the lepton pairs, $m_i$; $N$ denotes
the total number of events observed in the mass window for which the
likelihood is evaluated; $\mu_\mathtt{B}$ denotes the Poisson mean of
the total background yield; and $\mu=\mu_\mathtt{S}+ \mu_\mathtt{B}$
is the mean of the Poisson distribution from which $N$ is an
observation. The mass and width of the Breit--Wigner function are
denoted by $M$ and $\Gamma$, and $w$ denotes the width of the Gaussian
resolution function. The signal and background pdf's are denoted by
$f_\mathrm{S}$ and $f_\mathrm{B}$.

\subsection{Non-resonant searches}

The ADD and CI searches both use the same methodology and are based on
counting events above mass thresholds. In the ADD analysis an
optimal minimum mass threshold is chosen to maximize the limit on the
parameter $\LT$ using expected limits, and a similar procedure
is followed in the CI analysis, where the relevant parameter is the
energy scale $\LC$, and both destructive and constructive
interference scenarios are considered.

\subsection{Non-resonant likelihood function}

The probability of observing a number $N_{\text{obs}}$ of events is
given by the Poisson likelihood

\begin{equation}
\mathcal P \left( N_{\text{obs}} \right) = \frac{a^{N_{\text{obs}}}}{N_{\text{obs}}!} \re^{-a},
\end{equation}
where $a$ is the assumed Poisson mean. Both the background and a
potential signal can contribute to the Poisson mean, $a$, which can be
expressed as
\begin{equation}
a = (\epsilon_{\mathrm{s}} \sigma_{\mathrm{s}} + \epsilon_{\mathrm{b}}
\sigma_{\mathrm{b}}) \frac{N_{\text{obs,Z}}}{\sigma_{\cPZ}  \epsilon_{\cPZ}},
\label{eq:stat_model}
\end{equation}
where $\sigma_{s}$ and $\sigma_{\mathrm{b}}$ are the respective cross
sections of the signal and the background, and
$\epsilon_{s},\epsilon_{\mathrm{b}}$ are the total efficiencies
(including acceptances) for the signal and background,
respectively. The expected number of background events is estimated
using the number of events in a mass window of $\pm$30\GeV around the
$\cPZ$ peak, as in the resonance analysis. In Eq.~(\ref{eq:stat_model})
the relevant quantities are indicated by the subscript \cPZ.

\subsection{Uncertainties \label{sec:uncertanties}}

The sources of uncertainty are the same in all of the interpretations
of the observed spectra. However, because the $\zp$ analysis makes use
only of the background shape and the ADD and CI analyses are based on
counting events, the importance of the uncertainties differs.  In
general, the uncertainties have little effect on the derived limits,
particularly when these limits are set using regions where no events
are observed in the data.

The dominant uncertainty in the \zp analysis is from the determination
of $R_\epsilon$, the ratio of selection efficiency times detector
acceptance for the $\zp$ decay to that of the $\cPZ$ decay.  This
uncertainty is 3\% for the dimuon channel, 4\% for the dielectron
barrel-barrel channel, and 6\% for the barrel-endcap channel.  These
values reflect the uncertainty in the estimation of the detector
acceptance (including the contribution associated with the choice of
PDF set) and in the evaluation of the reconstruction efficiency,
particularly in the ``turn-on'' region at low mass.

In the dimuon channel, the effects of the uncertainties in the muon
transverse momentum resolution and in the transverse momentum scale at
high \pt are evaluated in the context of both the resonance and
non-resonant searches.  Different misalignment scenarios where
displacements can be incoherent or coherent are considered.  The
transverse momentum resolution uncertainty has a negligible effect in
the resonance analysis and leads to a 5\% uncertainty in the predicted
background in the non-resonant analyses.  Misalignments also lead to a
transverse momentum scale uncertainty of 5\% per \TeVns{}.  This has a
negligible effect on the resonance analysis, but does dominate the
yield uncertainty in the non-resonant search analyses.  The
uncertainty depends on the lower mass selection threshold $\mminll$,
and the rapidly falling spectrum leads to an increasing relative
uncertainty with the mass threshold.  In the ADD analysis $\mminll$ =
1900\GeV and the uncertainty is 41\%. Similarly, in the CI analysis
for destructive interference $\mminll = 1500\GeV$ resulting in an
uncertainty of 28\% and for constructive interference $\mminll =
1200\GeV$, giving a 20\% uncertainty.

In both channels the residual background from jets
misidentified as leptons is very small, and the uncertainty in this
background has a negligible effect on the limit determination.

The PDF uncertainty in the Drell--Yan cross section is evaluated using
the PDF4LHC procedure~\cite{PDF4LHC2,PDF4LHC}.  Cross sections are
calculated at next-to-next-to-leading order (NNLO) using
{\FEWZ} \cite{Gavin:2010az}. The NNLO PDF sets
MSTW08 \cite{Martin2009}, CT10 \cite{Gao2014}, and
NNPDF21 \cite{Ball:2011uy} implemented in LHAPDF \cite{LHAPDF} are
used to evaluate the PDF uncertainty as a function of mass.  The
resulting uncertainties can be parametrized as a quadratic function of
mass: ($2.76 + 3.03 \times 10^{-3} m + 2.38 \times 10^{-6}
m^{2}$)\% for dimuons and ($4.15 + 1.83 \times 10^{-3} m + 2.68 \times 10^{-6} m^{2}$)\%
for dielectrons, where $m$ is expressed in units of \GeVns{}.  These
uncertainties are included as a function of mass in the limit
calculations for all of the analyses. In the non-resonant dielectron
analyses the PDF uncertainty dominates.

The \POWHEG Monte Carlo simulation includes QCD effects at NLO and
electroweak effects at leading order (LO). The {\textsc{fewz}} and
{\textsc{horace}}~\cite{HORACE1,HORACE2,HORACE3,HORACE4,HORACE5,HORACE6}
programs are used to evaluate the effects of NLO electroweak
corrections and the addition of photon induced interactions.  The
overall correction is found to be approximately constant as a function
of the invariant mass, though the size of the photon induced correction
is rather sensitive to the PDF set used \cite{PIPDFRef}.  In the ADD and
CI analysis the correction and its uncertainty, namely $1.0\pm0.1$, is
used for the background prediction. Using {\FEWZ}, a K-factor
of $1.024\pm0.030$ is found for QCD NLO to NNLO processes. For the
$\zp$ analysis neither of these constant K-factors has any effect on
the background uncertainty because the function used to parametrize
the background is normalized to the data.

Common systematic uncertainties are taken to be fully correlated in
the calculation of combined limits.

\subsection{Exclusion limits}

A Bayesian limit setting procedure is used for all interpretations of
the observed mass spectra. A positive, flat prior is used for the signal cross
section as described in Refs.~\cite{CMS-dilep-2012,CMS-dilep-2010} and
log-normal priors for the systematic uncertainties. The Markov
Chain Monte Carlo approach is used for integration.

A prior flat in the signal cross section $\sigma_s$ yields excellent
frequentist coverage properties.  Previous searches~\cite{CMS2012:ADDdilepton,CMS_ci_dimuon_7_TEV}
presented results
with prior functions effectively flat in $1/\sqrt{\sigma_s}$, in
search regions where interference is negligible for the ADD and CI
models~\cite{ATLASCIADD2014,PhysRevD.87.015010}. Such priors can exclude a larger
$\MS$ (\LT) range, however, they are known to have a
frequentist coverage of less than 90\% for a 95\% Bayesian confidence
interval in non-resonant searches.

In the $\zp$ search analysis, only events in a window of $\pm$6 times
the mass resolution are considered in the limit setting procedure.  To
ensure the background is properly constrained, the lower edge of the
window is adjusted so that there are at least 400 events in the
mass window. The observed limits have been found to be robust and do
not significantly change for reasonable variations in the limit
setting procedure, such as modifications in the window for accepting events
in the likelihood and changes in the background normalization and
shape.

\begin{figure}
\centering
\includegraphics[width=\cmsFigWidth]{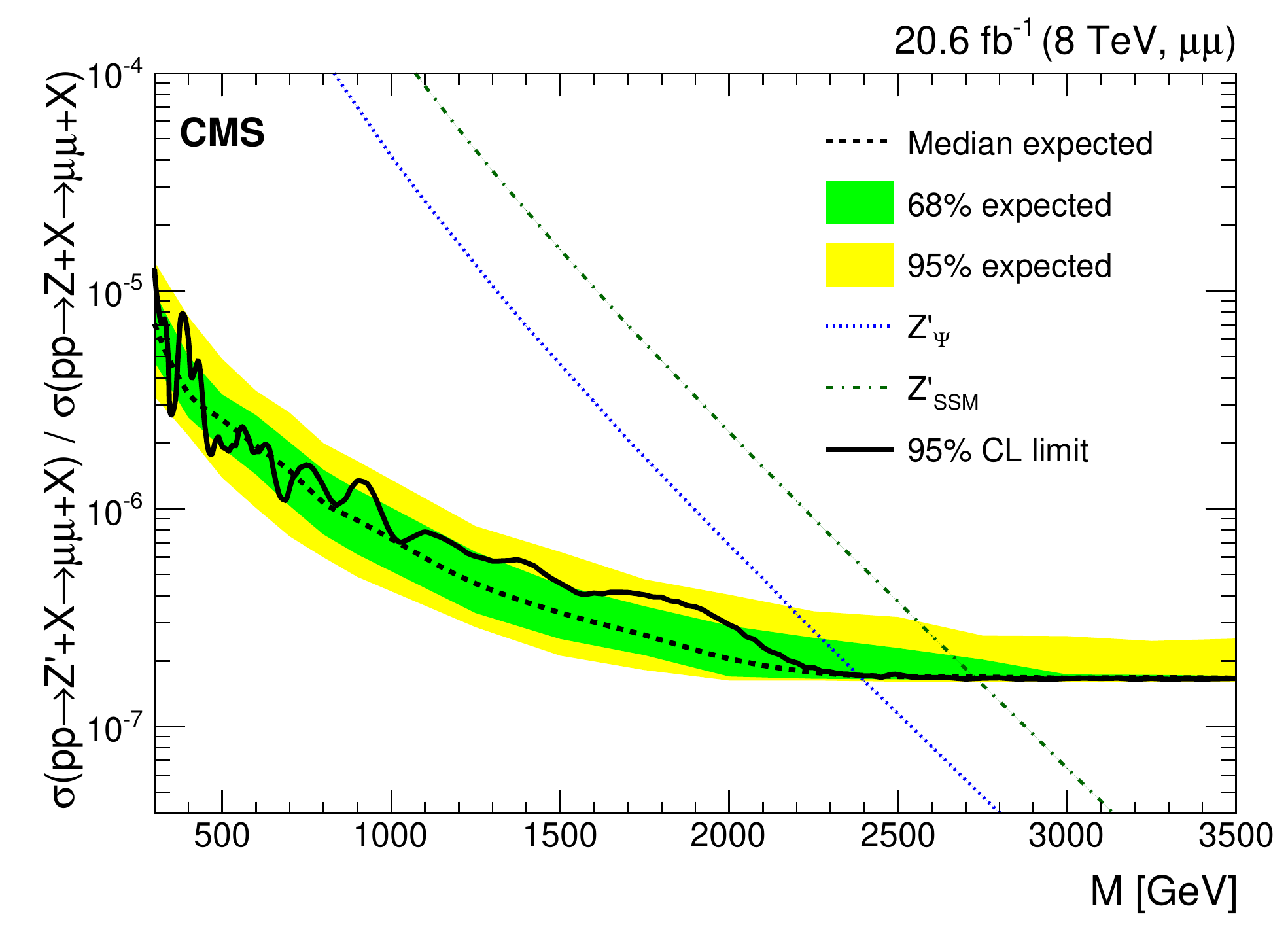}
\includegraphics[width=\cmsFigWidth]{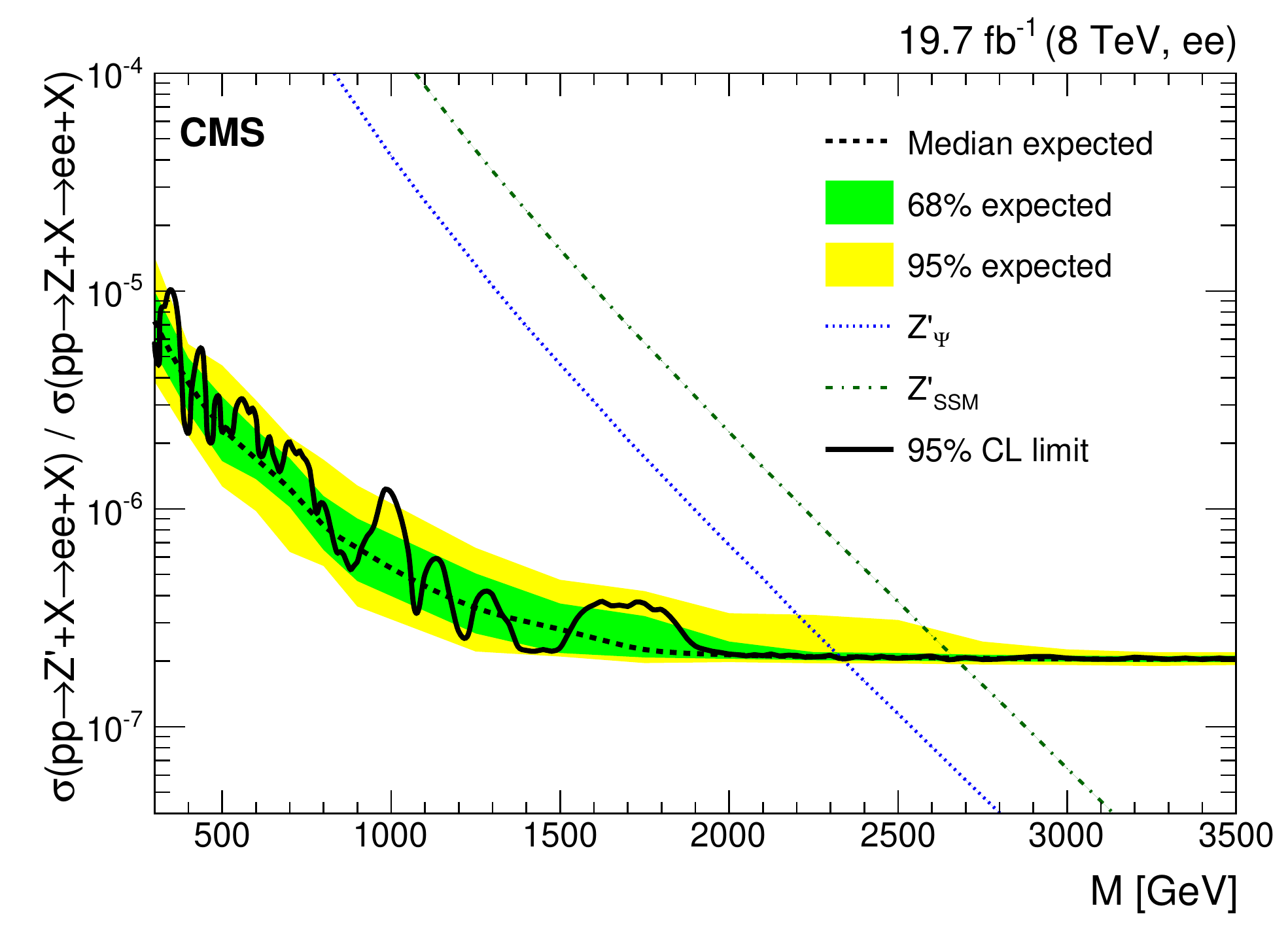}
 \caption{Upper limits as a function of the resonance mass $M$ on the
ratio of the product of production cross section and branching
fraction into lepton pairs for a spin-1 resonance relative to that of
$\cPZ$ bosons.  The limits are shown for dimuon (top) and dielectron
(bottom) final states.  The shaded bands correspond to the 68\% and
95\% quantiles for the expected limits.  Theoretical predictions for
spin-1 resonances, $\ZPSSM$ and $\ZPPSI$, are shown for comparison.}
\label{fig:limits}
\end{figure}

\begin{figure}
\centering
\includegraphics[width=\cmsFigWidth]{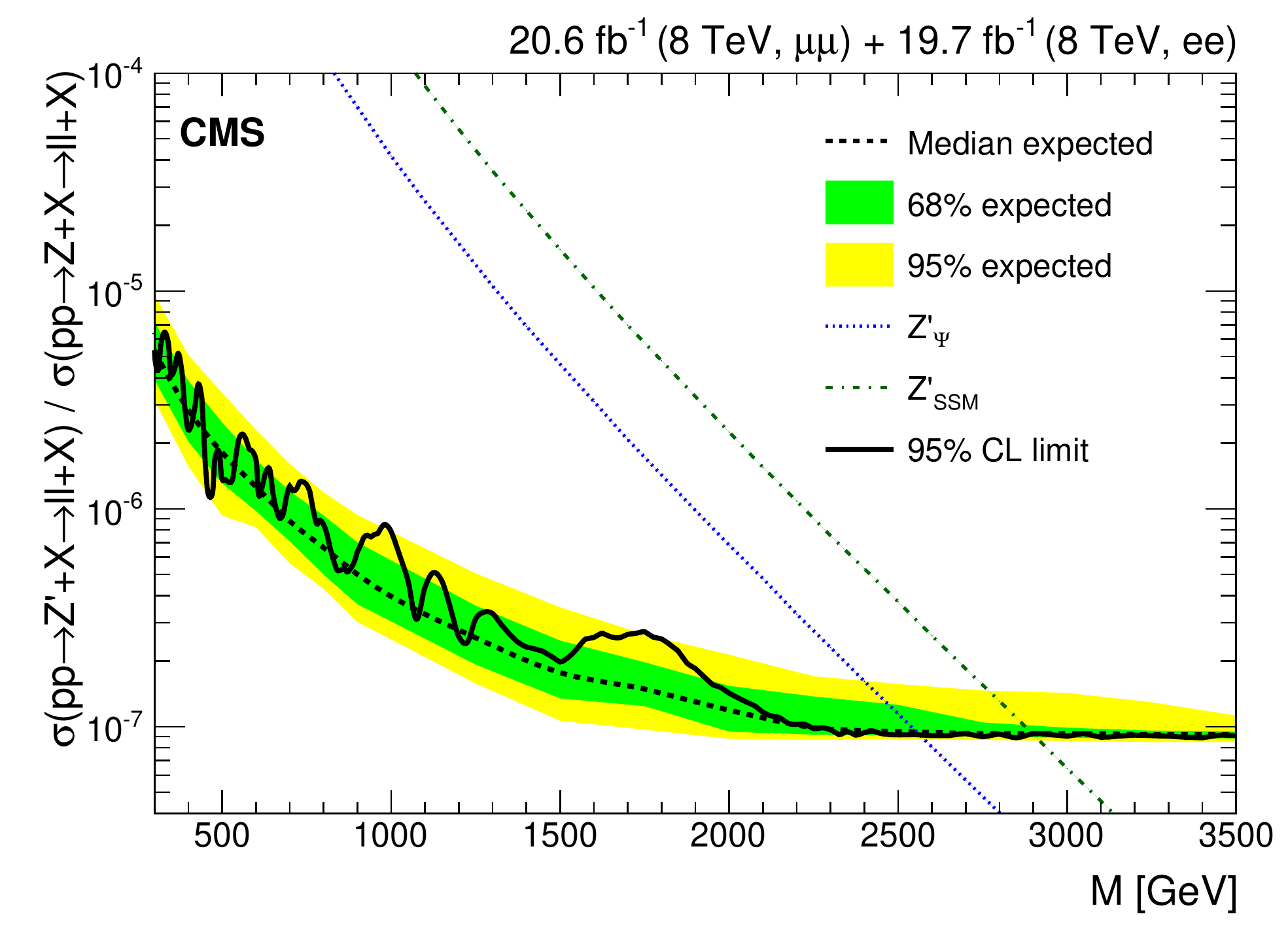}
\includegraphics[width=\cmsFigWidth]{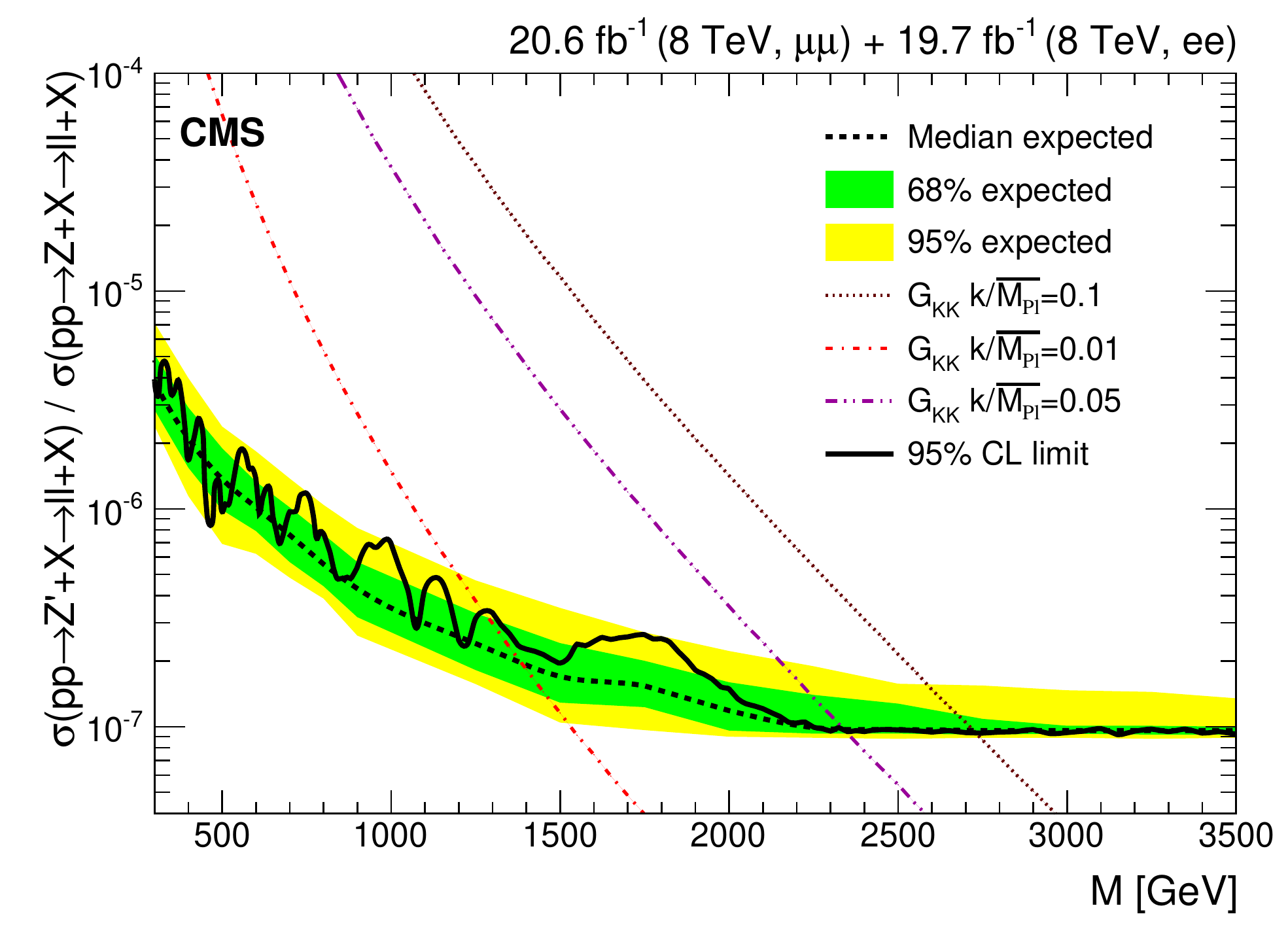}
 \caption{ Upper limits as a function of the resonance mass $M$ on the
   ratio of the product of cross section and branching fraction into
   lepton pairs relative to that of $\cPZ$ bosons, for final-state
   spins of 1 (top) and 2 (bottom). The shaded bands correspond to the
   68\% and 95\% quantiles for the expected limits. Theoretical
   predictions for spin-1 resonances, $\ZPSSM$ and $\ZPPSI$, and
   spin-2 RS gravitons are shown for comparison.
}
\label{fig:limits-comb}
\end{figure}

Figure~\ref{fig:limits} shows the observed and expected upper limits
on the ratio $R_{\sigma}$ of the production cross section times
branching fraction of a $\zp$ boson relative to that for a $\cPZ$
boson, for the dimuon and dielectron
channels. Figure~\ref{fig:limits-comb} shows the upper limits for the
combined dimuon and dielectron channels for the two spin hypotheses.
The figures also show the predicted ratios of cross section times
branching fraction for $\ZPSSM$ and $\ZPPSI$ production; together with
those for $\GKK$ production, with the dimensionless graviton coupling
to SM fields $k/\overline{M}_\mathrm{Pl}=0.05$, 0.01, and 0.1.  The LO
cross section predictions for $\ZPSSM$ and $\ZPPSI$ from \PYTHIA using
the CTEQ6.1 PDF set~\cite{CTEQ6L1} are corrected by a mass-dependent
K-factor obtained using
\textsc{zwprodp}~\cite{Accomando:2010fz,Hamberg:1990np,
vanNeerven:1991gh}, to account for the NNLO QCD contributions.
The calculated $\cPZpr$ cross sections include generated dileptons
with masses only within ${\pm}5\%\sqrt{s}$ of the nominal resonance
mass, to enhance sensitivity to a narrow-width
resonance~\cite{Accomando:2013}.  The NNLO prediction used for the
$\cPZg$ production cross section in the mass window of 60 to 120\GeV
is 1.117\unit{nb}, which was calculated using
{\textsc{fewz}}~\cite{Gavin:2010az}. The theoretical uncertainty is
expected to be 4\%, based on 7\TeV studies where factorization and
renormalization scales were varied and an uncertainty from PDFs
included.  No uncertainties in cross sections for the various
theoretical $\zp$ models are included when determining the limits.

For the dimuon channel, the 95\% CL lower limit on the mass of a $\zp$
resonance is \limitmumuZssm\ (\limitmumuZpsi)\TeV for $\ZPSSM$
($\ZPPSI$) and for the dielectron channel it is \limiteeZssm\
(\limiteeZpsi)\TeV for $\ZPSSM$ ($\ZPPSI$).  For the combined dimuon
and dielectron channels, the 95\% CL lower limit on the mass of a
$\zp$ resonance is \limitZssm\ (\limitZpsi)\TeV.  Randall--Sundrum
Kaluza--Klein gravitons are excluded below \limitmumuGhigh,
\limitmumuGlow, and \limitmumuGadd\TeV for couplings of 0.10, 0.05,
and 0.01 in the dimuon channel and similarly \limiteeGhigh,
\limiteeGlow, and \limiteeGadd\TeV in the dielectron channel. The
combined limits are \limitGhigh, \limitGlow, and \limitGadd\TeV. The
only limit that differs from the expected value is that for the
Randall--Sundrum Kaluza--Klein graviton with a coupling of 0.01, where
the expectation is 1.38\TeV. The limits quoted above are a commonly
used set of benchmarks. However, the model-independent method used to
derive these limits enables them to be reinterpreted in a
straightforward way in the context of any model that is characterized
by a narrow spin-1 or spin-2 resonance.  The discussion that follows
illustrates the versatility of these results.

The cross section for charged lepton-pair production via a $\zp$
vector boson can, in the narrow-width approximation (NWA), be
expressed in terms of the quantity $c_u w_u + c_d
w_d$~\cite{Carena:2004xs,Accomando:2010fz}.  The parameters $c_u$ and
$c_d$ contain information from the model-dependent $\zp$ couplings to
fermions in the annihilation of charge 2/3 and charge $-$1/3 quarks;
$w_u$ and $w_d$ contain the information about PDFs for the respective
annihilation at a given $\zp$ mass.
The translation of the experimental limits into the ($c_u$,$c_d$)
plane has been studied in the context both of the NWA and taking finite widths into account.  The procedures
have been shown to give the same results \cite{Accomando:2010fz}. A
further study including the effects of
interference~\cite{Accomando:2013} has demonstrated that with an
appropriate choice of the invariant mass window within which the cross
section is calculated, this approach can still be used.

\begin{figure}[htb]
\centering
 \includegraphics[width=\cmsFigWidth]{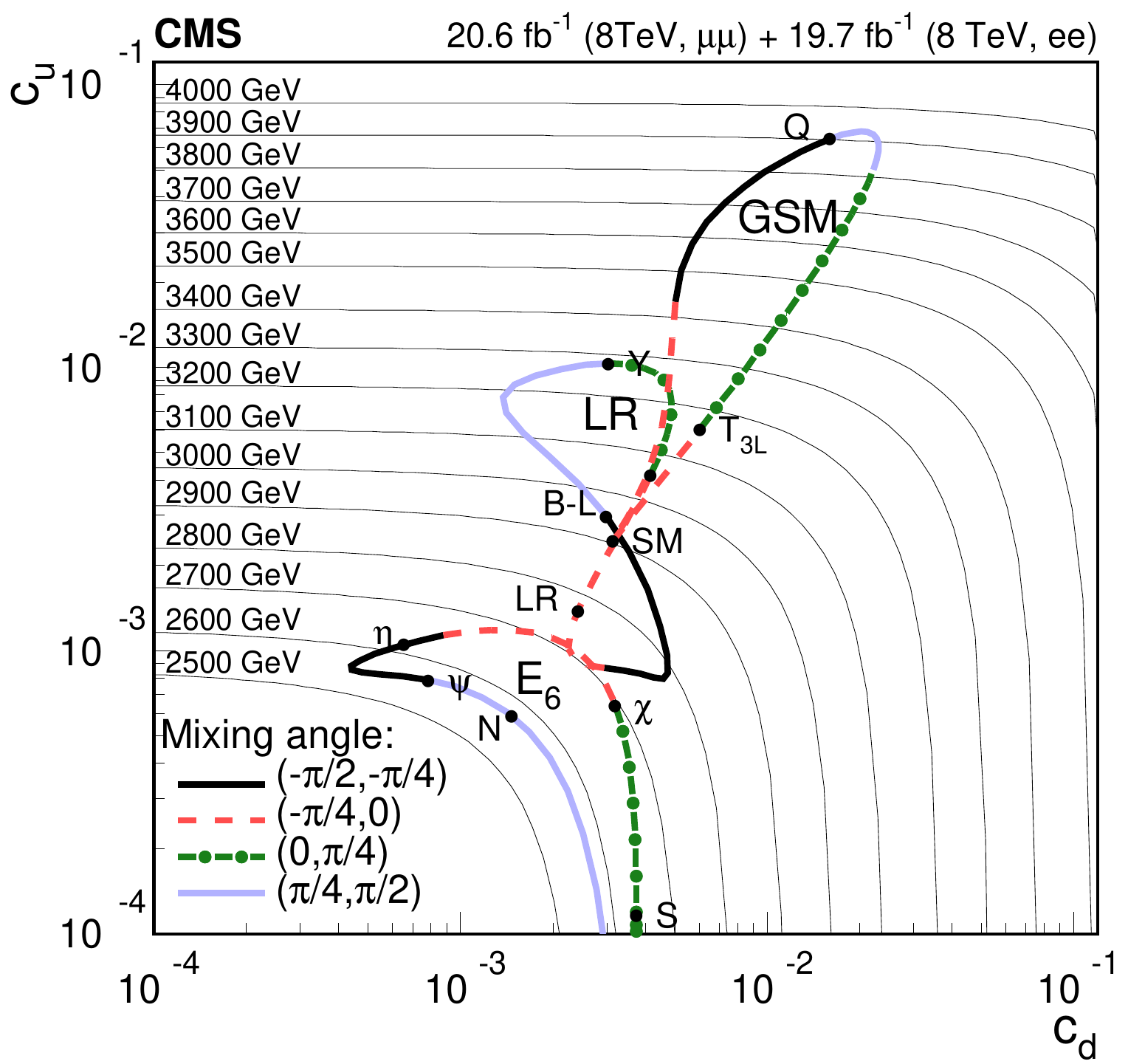}

\caption{ The combined limit on $\zp$ boson mass
for dimuon and dielectron channels at 95\% CL shown in the ($c_u$,$c_d$)
plane. The solid thin black lines represent the experimental upper limits
on ($c_u$,$c_d$) at
the masses specified in the figure. The contours representing the
GSM, Left-Right, and E$_6$ model classes are composed of thick line
segments. As indicated in the legend, the segment line styles
correspond to ranges of the mixing angle.}

\label{fig:CuCd}
\end{figure}

In Fig.~\ref{fig:CuCd} the limits on the $\zp$ mass are shown as lines
in the $(c_d,c_u)$ plane intersected by curves showing $(c_d,c_u)$ as
a function of a mixing parameter for various models. In this plane,
the thin solid lines labelled by mass are contours of cross
section with constant $c_u + (w_d/w_u)c_d$, where $w_d/w_u$ is in the
range 0.5--0.6 for the results relevant here.
In Ref.~\cite{Accomando:2010fz} a number of classes of models were defined,
which are illustrated here in Fig.~\ref{fig:CuCd}.  The Generalized
Sequential Model (GSM) class, where the generators of
$U(1)_{T_{3L}}$ and $U(1)_Q$ gauge groups~\cite{Accomando:2010fz} mix
with mixing angle $\alpha$,
includes the SM-like $\zp$ boson where the mixing angle is
$\alpha= -0.072 \pi$. The angles $\alpha =0$ and $\alpha=\pi/2$ define
the $T_{3L}$ and $Q$ benchmarks, respectively.
Also shown are contours for the E$_6$ model (with $\chi$, $\psi$,
$\eta$, $S$, and $N$ corresponding to angles 0, 0.5$\pi$, $-0.29\pi$,
0.13$\pi$, and 0.42$\pi$) and generalized LR models
(with $R$, $B-L$, $LR$, and $Y$ corresponding to angles 0, 0.5$\pi$,
$-0.13\pi$, and 0.25$\pi$)~\cite{Accomando:2010fz} .

The lower mass thresholds $\mminll$ that are used for setting limits
in the model of large extra dimensions are chosen to give the largest
expected limits.  In general, $\mminll$ depends on $\Mmax$,
the scale up to which the theory is valid as described in the
Introduction. Thus the value of $\mminll$ increases with increasing
$\Mmax$ until a plateau is reached for values of
$\Mmax$ above about 3\TeV. The optimal value of $\mminll$
is found to be 1.9\TeV for dimuons and 1.8\TeV for dielectrons.

\begin{figure}[htbp]
  \centering
    \includegraphics[width=0.49\textwidth]{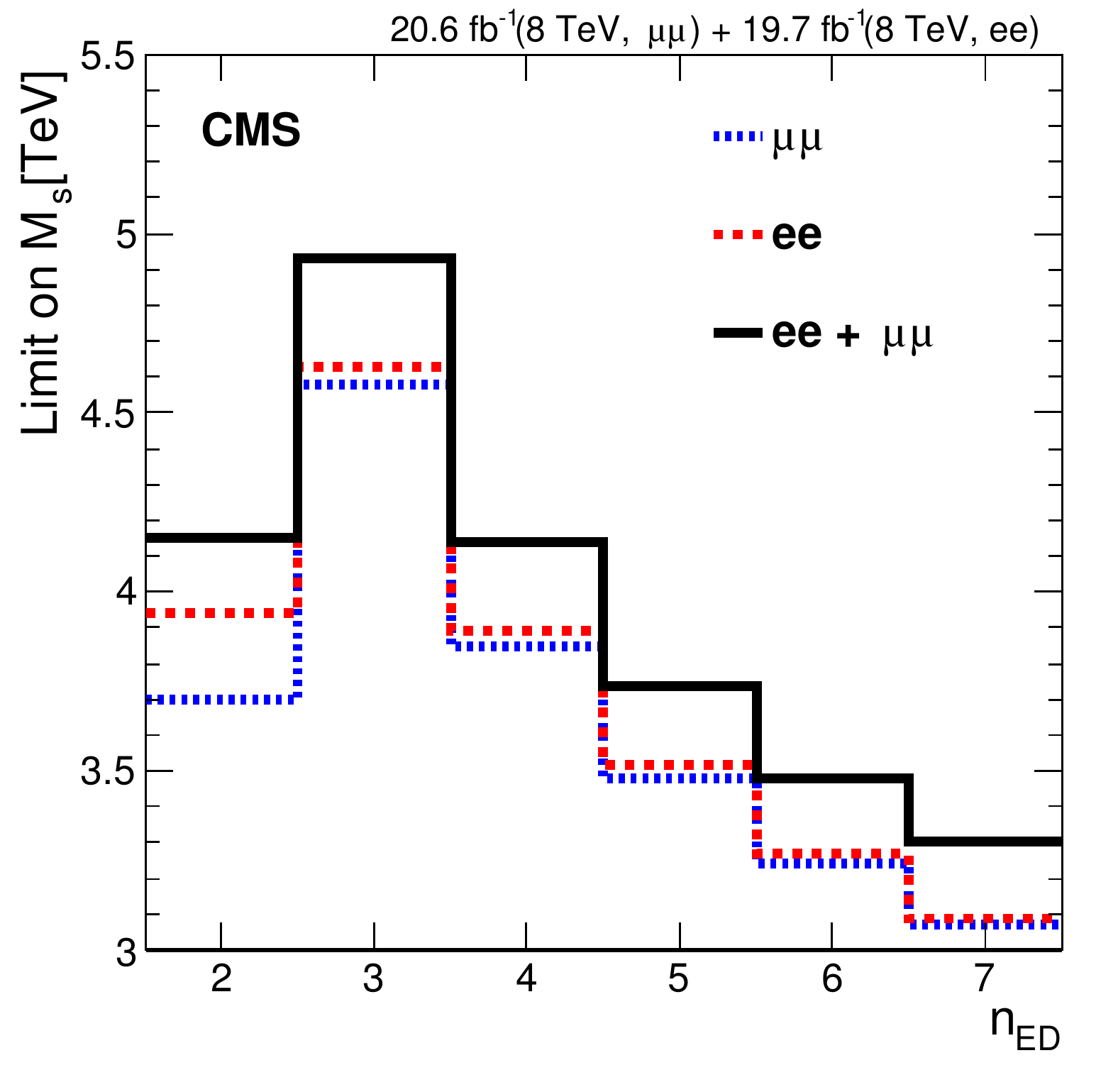}
  \caption{Observed 95\% CL limits on $M_{\mathrm{S}}$ for the dimuon (blue),
    dielectron (red) and combined (black) channel as a function of the
    number of extra spatial dimensions $n_{\mathrm{ED}}$.}
  \label{fig:ADD_limitplot_exp_results}
\end{figure}

Limits are set using the dimuon and dielectron mass spectra both
separately and combined. Above the respective optimal values of
\mminll the acceptance is 0.94 for both dimuon and dielectron channels
and the cross section limit is found to be 0.19 (0.18)\unit{fb} in the
dimuon (dielectron) channel.  Above a mass of 2.0\TeV the cross
section for both channels combined is 0.09\unit{fb}. The resultant
expected and observed limits on the ADD model parameters within the
GRW and HLZ conventions are shown in Table~\ref{tab:ADD-results}, and
the observed limits are shown in
Fig.~\ref{fig:ADD_limitplot_exp_results}.

\begin{table*}[htbp]
\centering
\topcaption{\label{tab:ADD-results}Observed and expected 95\% CL lower limits for the ADD model in the dilepton channels and the combination at 95\% CL within GRW and HLZ conventions for truncation at $\Mmax = \MS$ [HLZ] or $\Mmax = \LT$ [GRW].}
\begin{tabular}{ccccccccc}

ADD      &Limit &  $\LT$ (\TeVns{}) & \multicolumn{6}{c}{$\MS$ (\TeVns{}) [HLZ] }           \\
\cline{4-9}
K-factor &      &  [GRW]                        & $n = 2$ & $n = 3$ & $n = 4$ & $n = 5$ & $n = 6$ & $n = 7$   \\
\hline\hline
\multicolumn{9}{c}{$\mu\mu$, $m_{\mu\mu}>1.9\TeV$, $\sigma_{\mathrm{s}}<0.19\unit{fb}$ (0.19\unit{fb} expected) at 95\% CL} \\
\hline
1.0 &\multirow{2}{*}{expected }& 3.71&3.46 &4.42 &3.71 &3.36 &3.12 &2.95 \\
1.3 &                          & 3.84&3.69 &4.57 &3.84 &3.47 &3.23 &3.06 \\
\hline
1.0 &\multirow{2}{*}{observed }& 3.72 &3.48 &4.43 &3.72 &3.36 &3.13 &2.96 \\
1.3 &                          & 3.85 &3.70 &4.58 &3.85 &3.48 &3.24 &3.07 \\
\hline \hline
\multicolumn{9}{c}{$\Pe\Pe$, $m_{\Pe\Pe}>1.8\TeV$, $ \sigma_{\mathrm{s}}<0.18\unit{fb}$ (0.19\unit{fb} expected) at  95\%  CL} \\
\hline
1.0 &\multirow{2}{*}{expected }& 3.75&3.73 &4.47 &3.75 &3.39 &3.16 &2.99 \\
1.3 &                         & 3.88&3.91 &4.61 &3.88 &3.50 &3.26 &3.08 \\
\hline
1.0 &\multirow{2}{*}{observed }& 3.77 &3.75 &4.48 &3.77 &3.40 &3.17 &3.00 \\
1.3 &                          & 3.89 &3.94 &4.63 &3.89 &3.52 &3.27 &3.09 \\
\hline \hline
\multicolumn{9}{c}{$\mu\mu$ and $\Pe\Pe$, $m_{\ell\ell}>2.0\TeV$, $ \sigma_{\mathrm{s}}<0.09\unit{fb}$ (0.10\unit{fb} expected) at 95\%  CL} \\
\hline
1.0 &\multirow{2}{*}{expected }& 3.99&3.88 &4.74 &3.99 &3.60 &3.35 &3.17 \\
1.3 &                          & 4.13&4.13 &4.91 &4.13 &3.73 &3.47 &3.28 \\
\hline
1.0 &\multirow{2}{*}{observed }& 4.00 &3.90 &4.75 &4.00 &3.61 &3.36 &3.18 \\
1.3 &                          & 4.14 &4.15 &4.93 &4.14 &3.74 &3.48 &3.30 \\
\end{tabular}
\label{tab:combined_results}
\end{table*}

\begin{figure*}[ptbh]
\centering
         \includegraphics[width=0.49\textwidth]{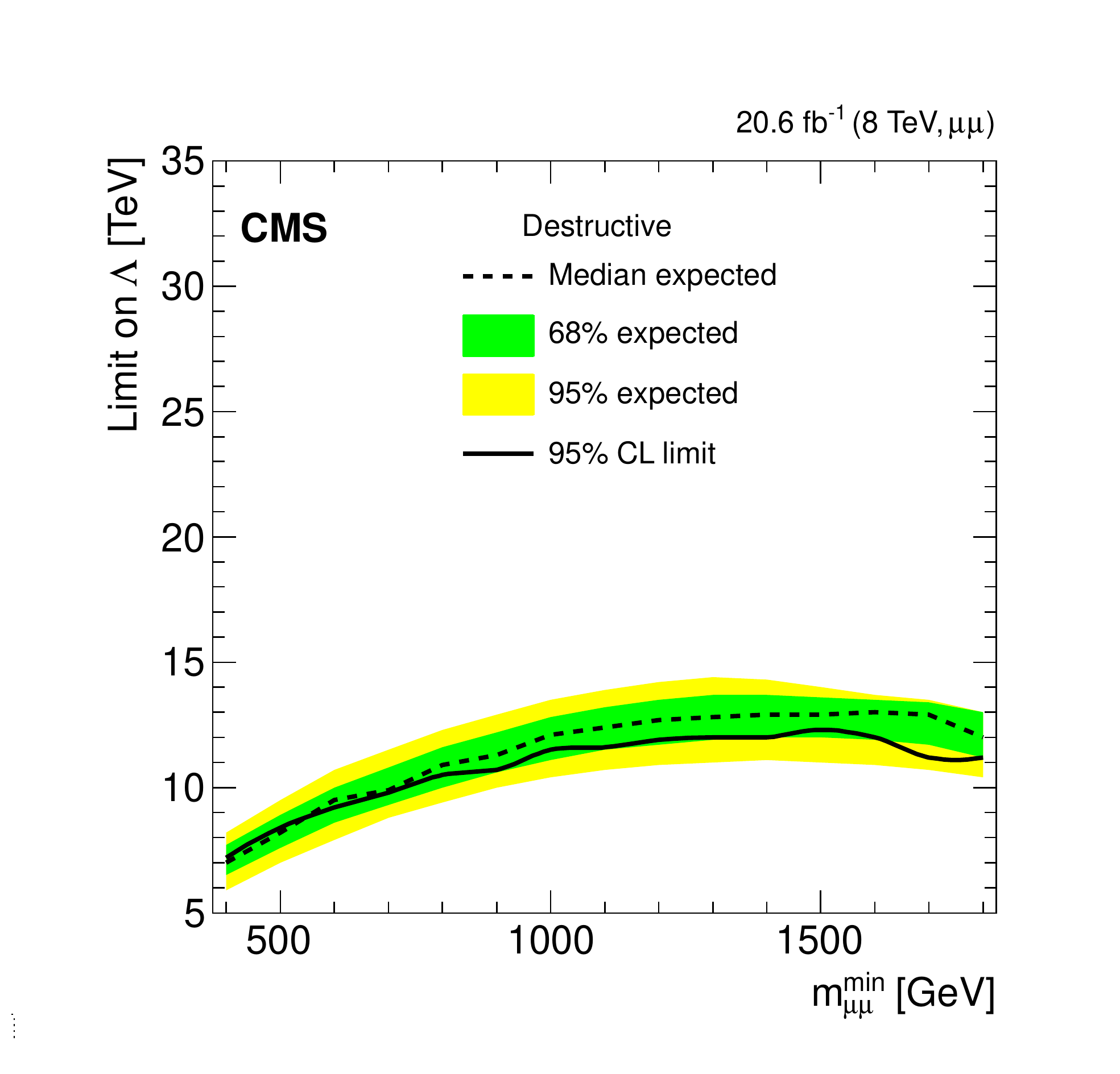}
         \includegraphics[width=0.49\textwidth]{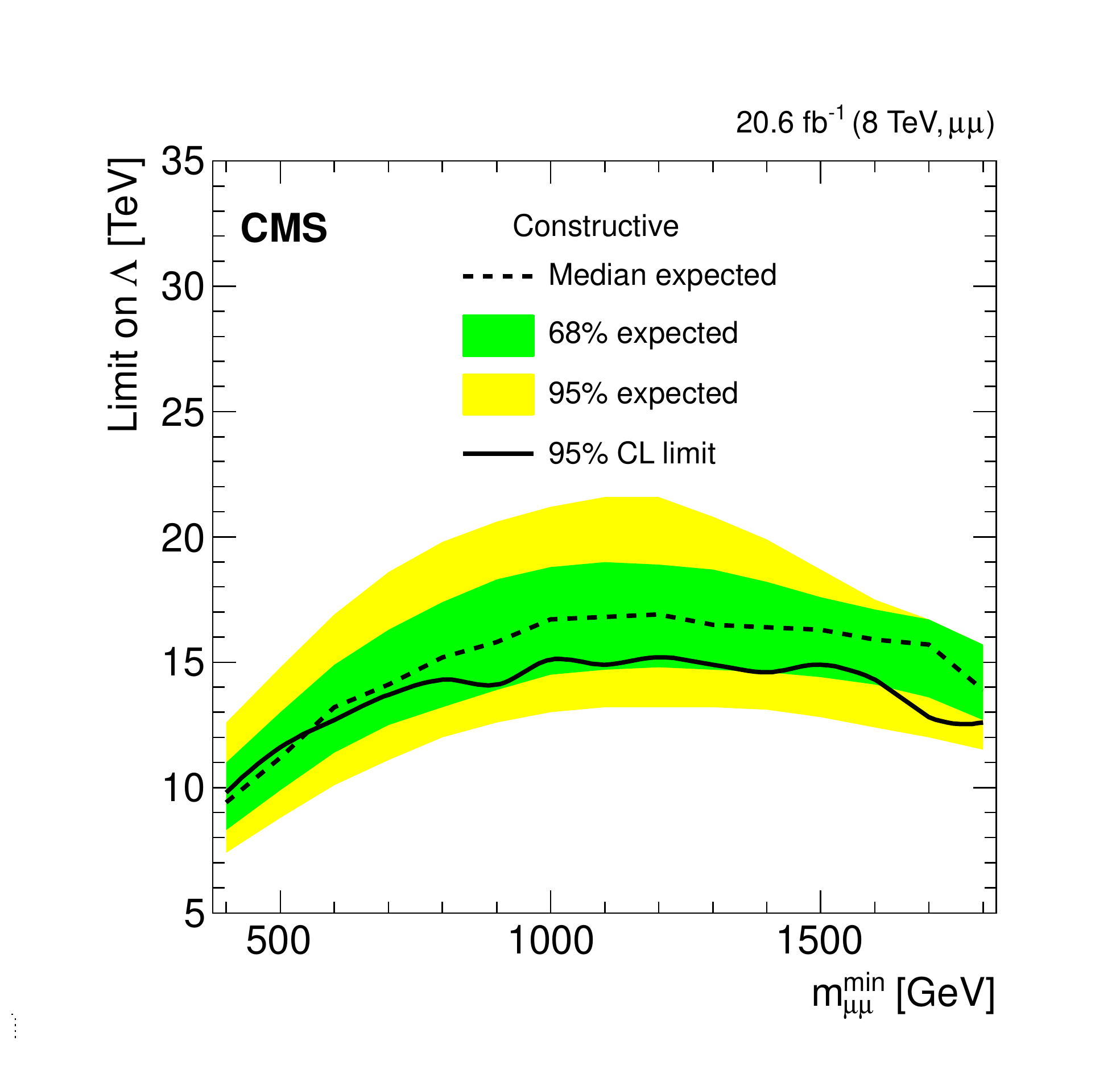}
         \includegraphics[width=0.49\textwidth]{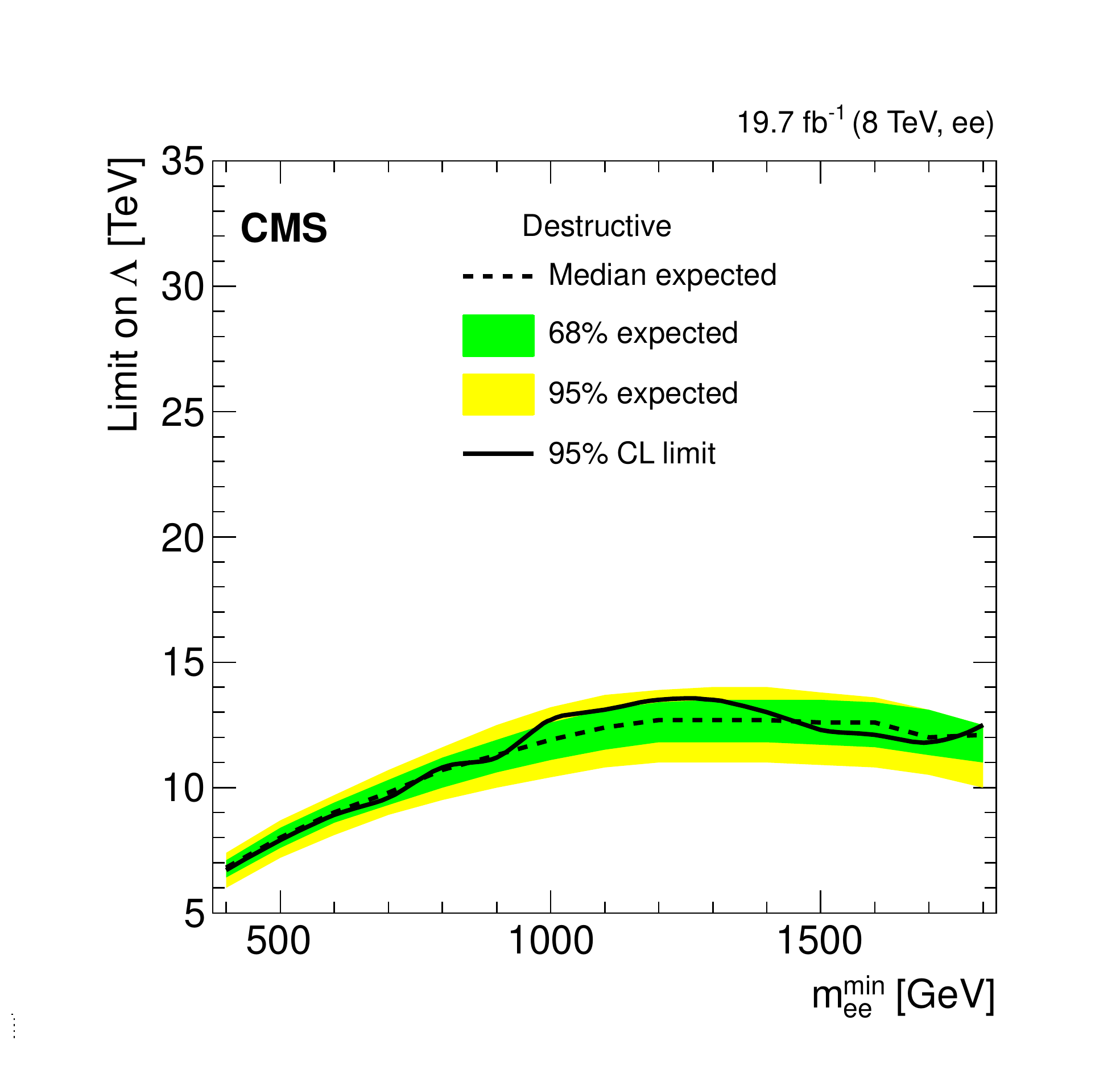}
         \includegraphics[width=0.49\textwidth]{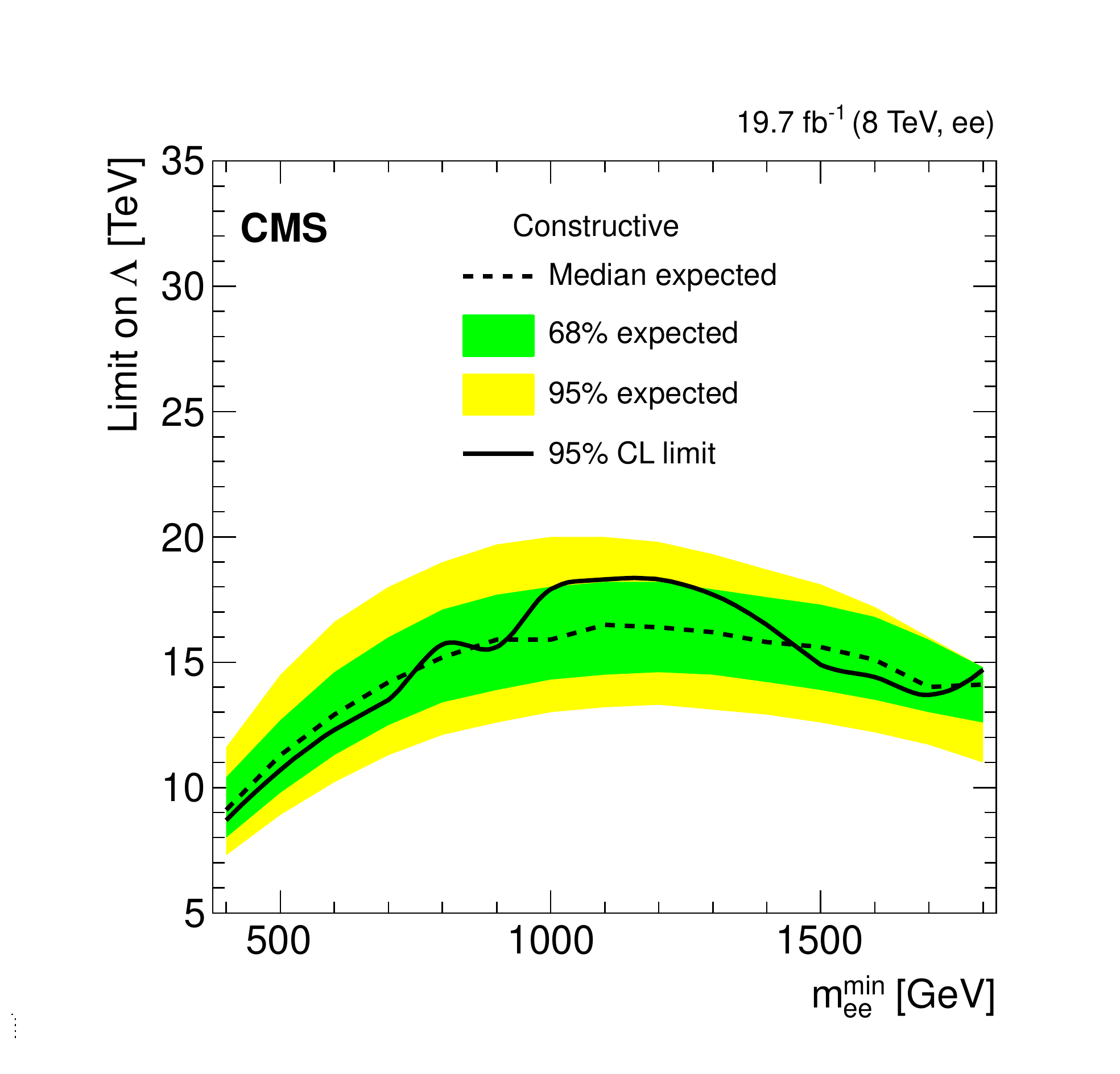}
\caption{\label{fig:dileptons_cl95}Observed and expected 95\% CL limits on \LC: (top) for the dimuon channel
      as a function of \mminmm for (left) destructive interference and (right)
      constructive interference; (bottom) for the  dielectron channel as
      a function of \mminee for (left) destructive interference and (right)
      constructive interference.}
\end{figure*}

In order to interpret the observed mass spectra in the context of the
contact interaction left-left isoscalar model described in the
Introduction, a similar procedure is followed. In this instance there
are two cases to consider, namely, where the interference is positive
or negative. In each of these cases the expected limits as a function
of a minimum mass $\mminll$ are found.  The value of $\mminll$ where
the expected limit is a maximum in the dimuon (dielectron) channel is
found to be 1500 (1300)\GeV for destructive interference and 1200 (1100)\GeV for constructive interference.  For these values of
$\mminll$ the observed (expected) limits on $\LC$ for destructive
and constructive interference respectively are: 12.0 (13.0)\TeV and
15.2 (16.9)\TeV for dimuons; and 13.5 (12.7)\TeV and 18.3 (16.5)\TeV
for dielectrons. The observed limits lie almost entirely within
1$\sigma$ of the expected value, as shown in
Fig.~\ref{fig:dileptons_cl95}.

Under the assumption that the contact interaction with quarks is
independent of lepton flavour, the dimuon and dielectron channels can
be combined by summing the yields in the two channels. Uncertainties
are added in quadrature, taking into account highly correlated sources
such as the PDF variations.
The observed (expected) limits for this combination are 13.1 (14.1)\TeV for destructive interference and 16.9 (17.1)\TeV for
constructive interference.

\section{Summary}

A search has been performed, using proton-proton collision data
collected at $\sqrt{s}=8\TeV$, for evidence of physics beyond the
standard model in dilepton mass spectra.  Data samples correspond to
integrated luminosities of $\anaLumimm$ and $\anaLumiee\fbinv$ for the
dimuon and dielectron channels, respectively. The spectra have been
found to be consistent with expectations from the standard model and
95\% confidence level limits have been set in the context of various
possible new physics models.

In the search for evidence of a narrow resonance, limits have been set
on the product of the cross section and branching fraction for new
boson production relative to the standard model $\cPZ$ boson
production.  Mass limits have been set on neutral gauge bosons as
follows: a $\zp$ with standard-model-like couplings has been excluded
below $\limitZssm\TeV$ and the superstring-inspired $\ZPPSI$ below
$\limitZpsi\TeV$; Randall--Sundrum Kaluza--Klein gravitons are
excluded below $\limitGhigh$, $\limitGlow$, and $\limitGadd\TeV$ for
couplings of 0.10, 0.05, and 0.01, respectively.  A notable feature of
the resonance analysis is that the limits may be reinterpreted in any
model predicting a resonance structure. To enable this, the limits
have been calculated in a model-independent way, and the
spin-dependent acceptance times efficiency as a function of mass has
been provided.

In the search for non-resonant deviations from standard model
expectations, limits have been set on the parameters within models of
extra dimensions and contact interactions.  Within the
Arkani-Hamed--Dimopoulos--Dvali model, lower limits have been set on
$\MS$, which characterizes the scale for the onset of quantum gravity.
These lower limits range from 4.9 to 3.3\TeV for 3 to 7 additional
spatial dimensions.  Within the context of the left-left isoscalar
contact interaction model, lower limits have been set on the energy
scale parameter $\LC$. For dimuons, the 95\% confidence level limit is
12.0 (15.2)\TeV for destructive (constructive)
interference. Similarly, for dielectrons the limit is 13.5 (18.3)\TeV
for destructive (constructive) interference.  The cross section limits
provided along with the dimuon and dielectron acceptances for the ADD
model considered may be used to reinterpret these results in the
context of other models resulting in non-resonant deviations from SM
expectations.

\begin{acknowledgments}
\hyphenation{Bundes-ministerium Forschungs-gemeinschaft Forschungs-zentren} We congratulate our colleagues in the CERN accelerator departments for the excellent performance of the LHC and thank the technical and administrative staffs at CERN and at other CMS institutes for their contributions to the success of the CMS effort. In addition, we gratefully acknowledge the computing centres and personnel of the Worldwide LHC Computing Grid for delivering so effectively the computing infrastructure essential to our analyses. Finally, we acknowledge the enduring support for the construction and operation of the LHC and the CMS detector provided by the following funding agencies: the Austrian Federal Ministry of Science, Research and Economy and the Austrian Science Fund; the Belgian Fonds de la Recherche Scientifique, and Fonds voor Wetenschappelijk Onderzoek; the Brazilian Funding Agencies (CNPq, CAPES, FAPERJ, and FAPESP); the Bulgarian Ministry of Education and Science; CERN; the Chinese Academy of Sciences, Ministry of Science and Technology, and National Natural Science Foundation of China; the Colombian Funding Agency (COLCIENCIAS); the Croatian Ministry of Science, Education and Sport, and the Croatian Science Foundation; the Research Promotion Foundation, Cyprus; the Ministry of Education and Research, Estonian Research Council via IUT23-4 and IUT23-6 and European Regional Development Fund, Estonia; the Academy of Finland, Finnish Ministry of Education and Culture, and Helsinki Institute of Physics; the Institut National de Physique Nucl\'eaire et de Physique des Particules~/~CNRS, and Commissariat \`a l'\'Energie Atomique et aux \'Energies Alternatives~/~CEA, France; the Bundesministerium f\"ur Bildung und Forschung, Deutsche Forschungsgemeinschaft, and Helmholtz-Gemeinschaft Deutscher Forschungszentren, Germany; the General Secretariat for Research and Technology, Greece; the National Scientific Research Foundation, and National Innovation Office, Hungary; the Department of Atomic Energy and the Department of Science and Technology, India; the Institute for Studies in Theoretical Physics and Mathematics, Iran; the Science Foundation, Ireland; the Istituto Nazionale di Fisica Nucleare, Italy; the Ministry of Science, ICT and Future Planning, and National Research Foundation (NRF), Republic of Korea; the Lithuanian Academy of Sciences; the Ministry of Education, and University of Malaya (Malaysia); the Mexican Funding Agencies (CINVESTAV, CONACYT, SEP, and UASLP-FAI); the Ministry of Business, Innovation and Employment, New Zealand; the Pakistan Atomic Energy Commission; the Ministry of Science and Higher Education and the National Science Centre, Poland; the Funda\c{c}\~ao para a Ci\^encia e a Tecnologia, Portugal; JINR, Dubna; the Ministry of Education and Science of the Russian Federation, the Federal Agency of Atomic Energy of the Russian Federation, Russian Academy of Sciences, and the Russian Foundation for Basic Research; the Ministry of Education, Science and Technological Development of Serbia; the Secretar\'{\i}a de Estado de Investigaci\'on, Desarrollo e Innovaci\'on and Programa Consolider-Ingenio 2010, Spain; the Swiss Funding Agencies (ETH Board, ETH Zurich, PSI, SNF, UniZH, Canton Zurich, and SER); the Ministry of Science and Technology, Taipei; the Thailand Center of Excellence in Physics, the Institute for the Promotion of Teaching Science and Technology of Thailand, Special Task Force for Activating Research and the National Science and Technology Development Agency of Thailand; the Scientific and Technical Research Council of Turkey, and Turkish Atomic Energy Authority; the National Academy of Sciences of Ukraine, and State Fund for Fundamental Researches, Ukraine; the Science and Technology Facilities Council, UK; the US Department of Energy, and the US National Science Foundation.

Individuals have received support from the Marie-Curie programme and the European Research Council and EPLANET (European Union); the Leventis Foundation; the A. P. Sloan Foundation; the Alexander von Humboldt Foundation; the Belgian Federal Science Policy Office; the Fonds pour la Formation \`a la Recherche dans l'Industrie et dans l'Agriculture (FRIA-Belgium); the Agentschap voor Innovatie door Wetenschap en Technologie (IWT-Belgium); the Ministry of Education, Youth and Sports (MEYS) of the Czech Republic; the Council of Science and Industrial Research, India; the HOMING PLUS programme of Foundation for Polish Science, cofinanced from European Union, Regional Development Fund; the Compagnia di San Paolo (Torino); the Consorzio per la Fisica (Trieste); MIUR project 20108T4XTM (Italy); the Thalis and Aristeia programmes cofinanced by EU-ESF and the Greek NSRF; and the National Priorities Research Program by Qatar National Research Fund.
\end{acknowledgments}
\bibliography{auto_generated}
\cleardoublepage \appendix\section{The CMS Collaboration \label{app:collab}}\begin{sloppypar}\hyphenpenalty=5000\widowpenalty=500\clubpenalty=5000\textbf{Yerevan Physics Institute,  Yerevan,  Armenia}\\*[0pt]
V.~Khachatryan, A.M.~Sirunyan, A.~Tumasyan
\vskip\cmsinstskip
\textbf{Institut f\"{u}r Hochenergiephysik der OeAW,  Wien,  Austria}\\*[0pt]
W.~Adam, T.~Bergauer, M.~Dragicevic, J.~Er\"{o}, C.~Fabjan\cmsAuthorMark{1}, M.~Friedl, R.~Fr\"{u}hwirth\cmsAuthorMark{1}, V.M.~Ghete, C.~Hartl, N.~H\"{o}rmann, J.~Hrubec, M.~Jeitler\cmsAuthorMark{1}, W.~Kiesenhofer, V.~Kn\"{u}nz, M.~Krammer\cmsAuthorMark{1}, I.~Kr\"{a}tschmer, D.~Liko, I.~Mikulec, D.~Rabady\cmsAuthorMark{2}, B.~Rahbaran, H.~Rohringer, R.~Sch\"{o}fbeck, J.~Strauss, A.~Taurok, W.~Treberer-Treberspurg, W.~Waltenberger, C.-E.~Wulz\cmsAuthorMark{1}
% ...
\vskip\cmsinstskip
% ...
1:~~Also at Vienna University of Technology, Vienna, Austria\\
2:~~Also at CERN, European Organization for Nuclear Research, Geneva, Switzerland\\
% ...

\end{sloppypar}
\end{document}